\documentclass{pasa}%

\title[Interstellar extinction in twenty open star clusters]{Interstellar extinction in twenty open star clusters}
\author[Rangwal et al.]{Geeta Rangwal$^{1}$\thanks{E-mail: geetarangwal91@gmail.com},
R. K. S. Yadav$^{2}$\thanks{E-mail:rkant@aries.res.in}, Alok Durgapal$^{1}$\thanks{E-mail:alokdurgapal@gmail.com}, D. Bisht$^{3}$\thanks{E-mail:dbisht@prl.res.in}\\
    \\
    $^{1}$ Center of Advanced Study, Department of Physics, D. S. B.
           Campus, Kumaun University Nainital 263002, India.\\
    $^{2}$ Aryabhatta Research Institute of Observational Sciences,
           Manora Peak, Nainital 263002, India.\\
    $^{3}$ Astronomy and Astrophysics Division, Physical Research Laboratory, Ahmedabad 380009, Gujarat, India.\\
    }
\jid{PASA}
\doi{10.1017/pas.\the\year.xxx}
\jyear{\the\year}

\begin{document}%
\begin{abstract}

The interstellar extinction law in twenty open star clusters namely Berkeley 7, 
Collinder 69, Hogg 10, NGC 2362, Czernik
43, NGC 6530, NGC 6871, Bochum 10, Haffner 18, IC 4996, NGC 2384, NGC 6193,
NGC 6618, NGC 7160, Collinder 232, Haffner 19, NGC 2401, NGC 6231, NGC 6823
and NGC 7380 have been studied in the optical and near-IR wavelength ranges. 
The difference between maximum and minimum values of 
$E(B-V)$ indicates the presence of non-uniform extinction in all the clusters except
Collinder 69, NGC 2362 and NGC 2384. The colour excess ratios are consistent with a
normal extinction 
law for the clusters NGC 6823, Haffner 18, Haffner 19, NGC 7160, NGC 6193, NGC 2401,
NGC 2384, NGC 6871, NGC 7380, Berkeley 7, Collinder 69 and IC 4996. We have found that 
the differential color-excess $\Delta E(B-V)$, which may be due to the occurrence of 
dust and gas inside the clusters, decreases with the age of the clusters. A spatial variation of
color excess is found in NGC 6193 in the sense that it decreases from east to
west in the cluster region. For the clusters Berkeley 7, NGC 7380 and NGC 6871,
a dependence of color excess $E(B-V)$ with spectral class and luminosity is
observed. Eight stars in Collinder 232, four stars in NGC 6530 and one star in
NGC 6231 have excess flux in near-IR. This indicates that these stars
may have circumstellar material around them.

\end{abstract}
\begin{keywords}
Star cluster: Reddening, interstellar dust-
  Interstellar extinction.
\end{keywords}
\maketitle%
\section{INTRODUCTION}
\label{sec:intro}

Interstellar dust is an important component of the interstellar medium. It is the remnant
of star formation and stellar evolution processes. Interstellar dust grains can transmit,
redirect and transmute the starlight (Fitzpatrick 2004). Due to this property of dust grain the actual
distances and magnitude determination of astronomical objects is often very difficult (Pandey et al. 2002).
For this reason it is very important to have knowledge of interstellar dust in the line of
sight of the objects under study. The interstellar extinction is caused by two distinct sources, a diffuse
component associated by the general interstellar medium and a variable component 
associated with localized region of higher mean density (Martin \& Whittet 1990, Turner
1994).
Studies of the variable component gives information about the composition of stars.
Extinction properties of dust are wavelength dependent, so investigation of
interstellar extinction law is the best way to determine the dust properties
(Yadav \& Sagar 2001, Fitzpatrick 2004). Young
open star clusters are the ideal objects for this kind of study because they may contain
gas and dust around early type ($O$, $B$ and $A$) stars (Hayashi 1970, Larson 1973, McNamara 1976,
Warner, Strom \& Strom 1979, Yadav \& Sagar 2001).

Trumpler studied the interstellar dust for the first time in 1920 and then
it was studied by many investigators. Krelowski and Strobel (1983) analysed interstellar
extinction law in two stellar
aggregates namely Sco OB2 and Per OB1 and found different extinction
curves for these two aggregates and their differences are large in far UV wave band.
They concluded that the differences in FUV (Far-Ultra-Violet) fluxes for these fields
is due to the
circumstellar dust and also confirmed the presence of two populations of grains.
Kiszkurno et al.\ (1984) investigated twenty large OB associations at different
galactic positions and found similar colour excess curves in the FUV range.
They also concluded that these complexes are obscured by same type of
interstellar material. Sagar (1987) studied interstellar
extinction in 15 open star clusters and found non-uniform extinction across 10 of the 
clusters. When the value of the inferred colour excess varies from one part of cluster region to another one, 
this is known as non-uniform extinction. The possible reasons behind the non-uniform extinction
has been discussed by many investigators. Sagar (1987) and Sagar \& Qian (1989) stated that this non-uniformity 
is due to the hot and ionised
circumstellar dust around the stars. Samson (1975), McCuskey \& Hauk (1964) and
Stone (1977) proposed that
dust shells around stars and local dust clouds lying in the cluster direction
is responsible for non-uniform extinction.
Tapia et al. (1988) studied the extinction law in clusters using 200 stars located within four open clusters
situated at same distance from the Sun. They found that these clusters have different
reddening values and that they follow the normal extinction law for $\lambda > 0.55~
\mu m$. They also interpreted the anomalies of the extinction law in then $U$ and $B$ bands because of the
presence of dense intra-cluster dust cloud of different grain properties in comparision with normal
dust grains. Cardelli et al. (1989) derived the mean extinction law
for the wavelength range $0.12~\mu m\leq\lambda\leq 3.5~\mu m$. This extinction law is applicable
for the diffuse as well as the dense interstellar medium and depends only on one parameter $R_{V}$, the
total to selective extinction ratio.

\begin{table}
\tiny
\centering
\caption{General information about the clusters under study taken from WEBDA website.
$N_{S}$ denotes the number of stars used in the analysis.}
\begin{tabular}{llrrccc}
\hline\hline
IAU number  & Clusters   &    $l$        &   $b$      &  Distance      &  log(age)    &     $N_{S}$       \\
            &                &  (deg)      & (deg)    &    (kpc)       &   (yr)   \\
\hline

C0150+621   &  Berkeley 7   &  130.13   & 0.37    &    2.57      &    6.60     &  40   \\
C0532+099   & Collinder 69  &  195.05   &-12.00   &    0.44      &    7.05     &  08   \\
C1108-601   &   Hogg 10     &  290.80   & 0.07    &    1.77      &    6.78     &  26   \\
C0716-248   &  NGC 2362     &  238.18   &-5.54    &    1.39      &    6.91     &  17   \\
C2323+610   & Czernik 43    &  112.84   & 0.13    &    2.50      &    7.70     &  24   \\
C1801-243   &  NGC 6530     &    6.08   &-1.33    &    1.33      &    6.87     &  33   \\
C2004+356   &  NGC 6871     &   72.64   & 2.08    &    1.57      &    7.00     &  83   \\
C1040-588   &  Bochum 10    &  287.03   &-0.32    &    2.03      &    6.85     &  11   \\
C0750-262   & Haffner 18    &  243.11   & 0.44    &    6.03      &    6.00     &  56   \\
C2014+374   &  IC 4996      &   75.36   & 1.31    &    1.73      &    6.95     &  21   \\
C0722-209   &  NGC 2384     &  235.39   &-2.41    &    2.12      &    6.00     &  10   \\
C1637-486   &  NGC 6193     &  336.70   &-1.57    &    1.15      &    6.77     &  20   \\
C1817-162   &  NGC 6618     &   15.09   &-0.74    &    1.30      &    6.00     &  11   \\
C2152+623   &  NGC 7160     &  104.01   & 6.45    &    0.79      &    7.28     &  14   \\
C1042-59   & Collinder 232  &  287.49   &-0.54    &    2.99      &    6.67     &  26   \\
C0750-261   &  Haffner 19   &  243.08   & 0.52    &    5.09      &    6.93     &  41   \\
C0727-138   &  NGC 2401     &  229.66   & 1.85    &    6.30      &    7.40     &  33   \\
C1650-417   &  NGC 6231     &  243.46   & 1.18    &    1.24      &    6.84     & 123   \\
C1941+231   &  NGC 6823     &   59.40   &-0.14    &    1.89      &    6.82     &  77   \\
C2245+578   &  NGC 7380     &  107.14   &-0.88    &    2.22      &    7.08     &  88   \\
\hline
\end{tabular}
\label{inf}
\end{table}

Pandey et al. (1990) found that differential extinction which is due to a non-uniform
distribution of star forming material
decreases systematically with the age of the cluster.
Yadav \& Sagar (2001) studied interstellar extinction law in 15 open star clusters and
found the presence of non-uniform extinction in all the clusters. They also found
that most of the clusters follow the normal extinction law for $\lambda < \lambda_{J}$, 
more specifically that the value of the colour excess for these clusters is close to the standard
value derived for the general interstellar medium in this waveband, but
show anomalous behaviour for $\lambda \geq \lambda_{J}$.
Pandey et al. (2002) studied 14 open clusters located around $l \sim 130^{\circ}$ and found
that dust along these objects follow the normal interstellar extinction law for $\lambda \geq \lambda_{J}$,
but for shorter wavelengths it depends upon $R_{cluster}$ (total-to-selective absorption
in the cluster region). Joshi (2005) studied the distribution of interstellar matter
near the Galactic plane on the basis of 722 open star clusters and found that $\sim$
90$\%$ of the absorbing material lies within $-5^{\circ} \leq b \leq 5^{\circ}$ of the
Galactic plane. Nishiyama et al. (2005) studied the interstellar extinction law in $J,H$
and $K_{S}$ bands towards the Galactic center in the region
$l \leq 2^{\circ}.0$ and $0^{\circ}.5 \leq b \leq 1^{\circ}.0$ using Red Clump (RC) stars and
concluded that the extinction law is not universal even in the infra-red
band.

In the light of above discussions, we can say that there are numerous studies available regarding
the interstellar extinction in the direction of different objects. But, yet there are number of
clusters for which less information is available about the distribution of dust and
gas in our Galaxy. In order to study the distribution of dust and gas in the Milky Way, we need
to consider different young open star clusters. 
Therefore, we have in this study
considered twenty open star clusters to study interstellar extinction law in the optical and near-IR
bands.

The structure of the article is as follows. Section 2 describes the data sources and cluster selection
criteria. Section 3 describes the data analysis while Section 4 represents the conclusions of the present
study.

\section{Cluster selection and data set} \label{sec:obs}

\begin{figure}
    \centering
    \includegraphics[width=9cm,height=7cm]{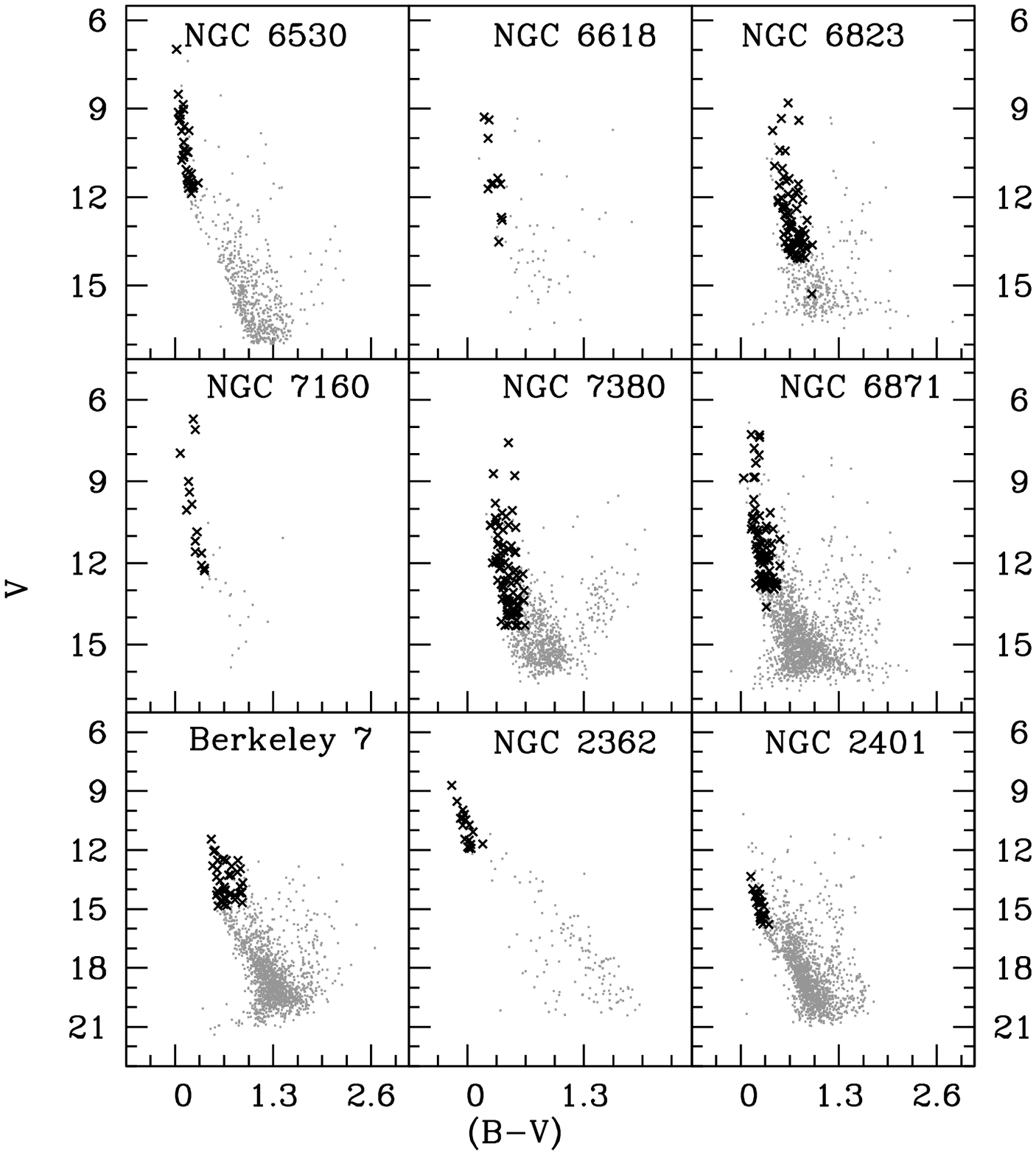}
    \includegraphics[width=9cm,height=7cm]{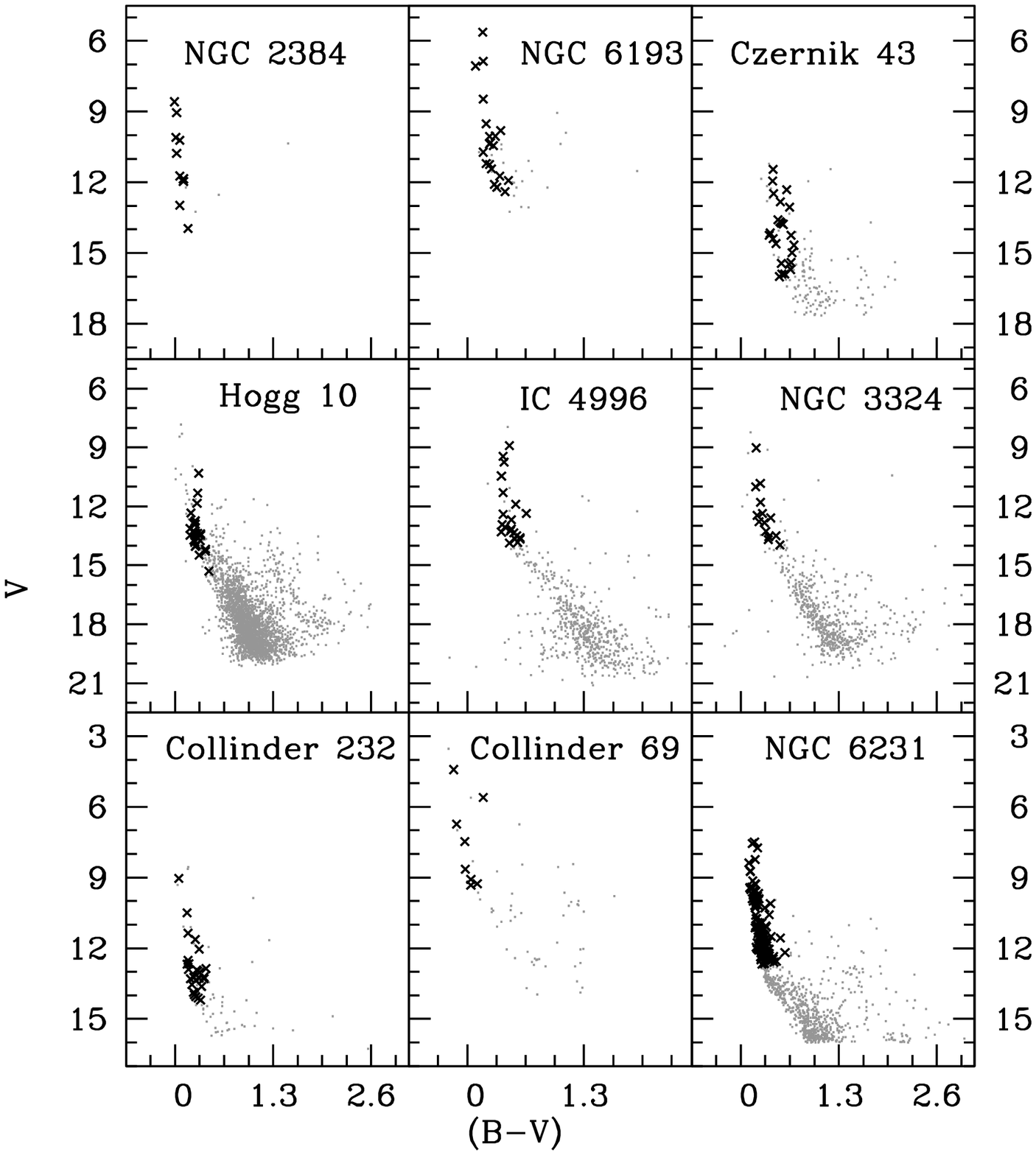}
    \includegraphics[width=6cm,height=4cm]{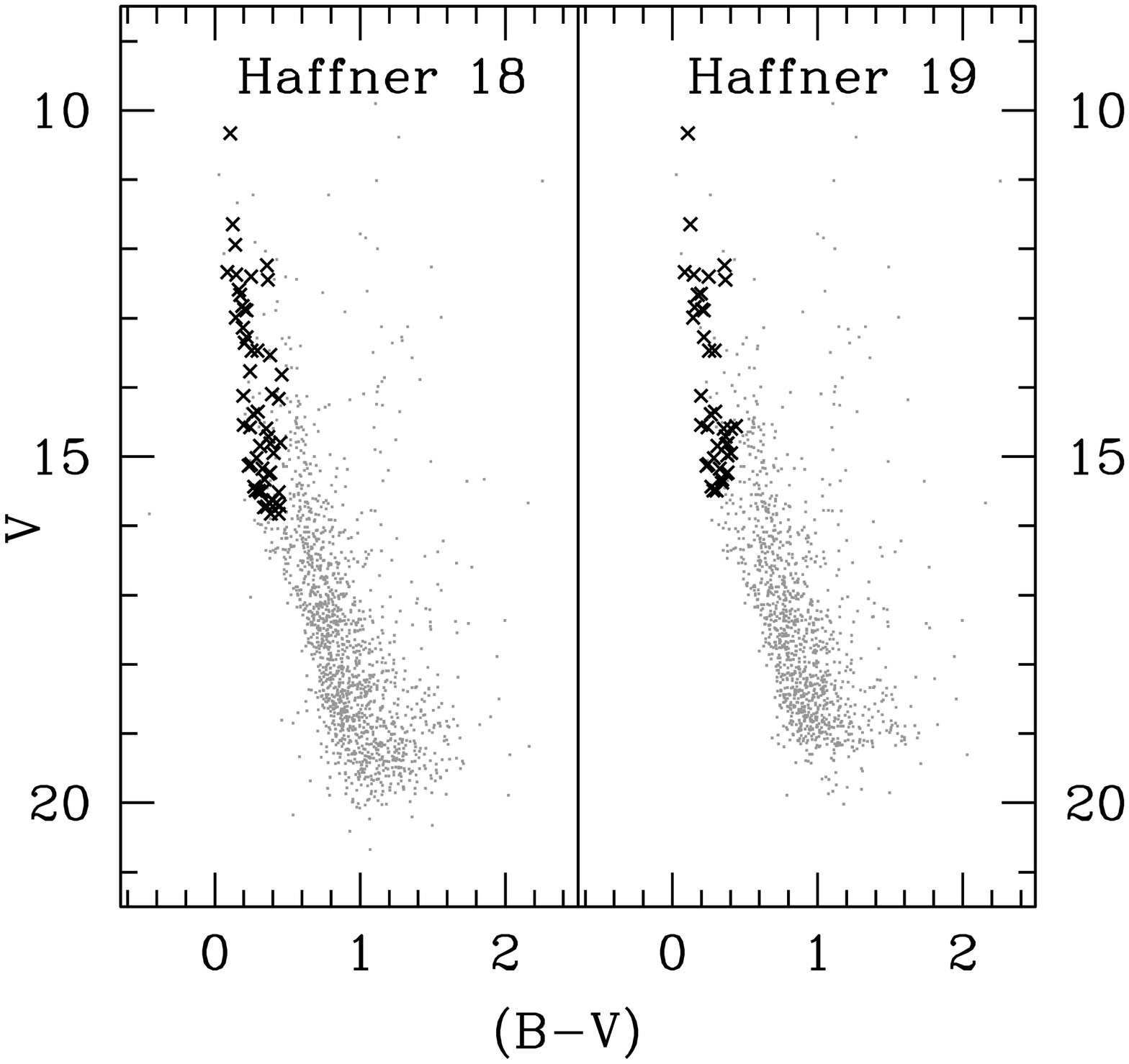}
    \caption{The $V$, $(B-V)$ colour-magnitude diagrams for all twenty clusters under study.
    The gray points represent all stars of the cluster and black crosses represent the stars 
    used for the present study.}
    \label{cmd}
  \end{figure}

To study the extinction properties in the clusters, we have selected
twenty young open star clusters from the WEBDA database, which have age younger than 50 Myr.
The selection criteria of these clusters is based on the availability of $UBVRI$ 
photometric data and previous studies on extinction. These 
clusters have been considered first time for interstellar extinction study.
These clusters have distances ranging from 0.44 kpc to 6.3 kpc. In young open star
clusters, the early type stars may have dust and gas around them. Therefore, for the
present analysis we have selected only early type stars ($O$, $B$ and $A$) in each cluster.
These stars have been selected using colour-colour and colour-magnitude diagrams (CMDs).
The CMDs of the clusters are shown in Fig. \ref{cmd}. The crosses represent the selected 
stars for the present analysis while gray points are the remaining stars of the clusters.
The $UBVRI$ data for these clusters are taken from the WEBDA (http://www.univie.ac.at/webda/) database
while $JHK$ data are taken from 2MASS (Two Micron All Sky Survey) catalogue.
These clusters are distributed along the galactic plane with longitude from $6^{\circ}.08$
to $336^{\circ}.70$. Almost all the clusters under study have $UBVRIJHK$ data. The optical data
set have error $\sim$ 0.01 mag in $V$ and $\sim$ 0.02 mag in $(B-V)$, $(V-R)$ and $(V-I)$ while
$\sim$ 0.03 mag in $(U-B)$. The $JHK$ data set have error $\sim$ 0.05 mag in $J$ mag. The spectroscopic
data is taken from the WEBDA database wherever available. Due to the unavailability of
spectroscopic data for many stars, photometric Q-method (Johnson \& Morgan 1953) have been
used. The maximum difference between the spectral type as identified by Q-method and that
of available spectral type data is about two spectral sub-classes. 
The general information about the selected open star clusters is listed in Table \ref{inf}.

\begin{figure}
    \centering
    \includegraphics[width=9cm,height=7cm]{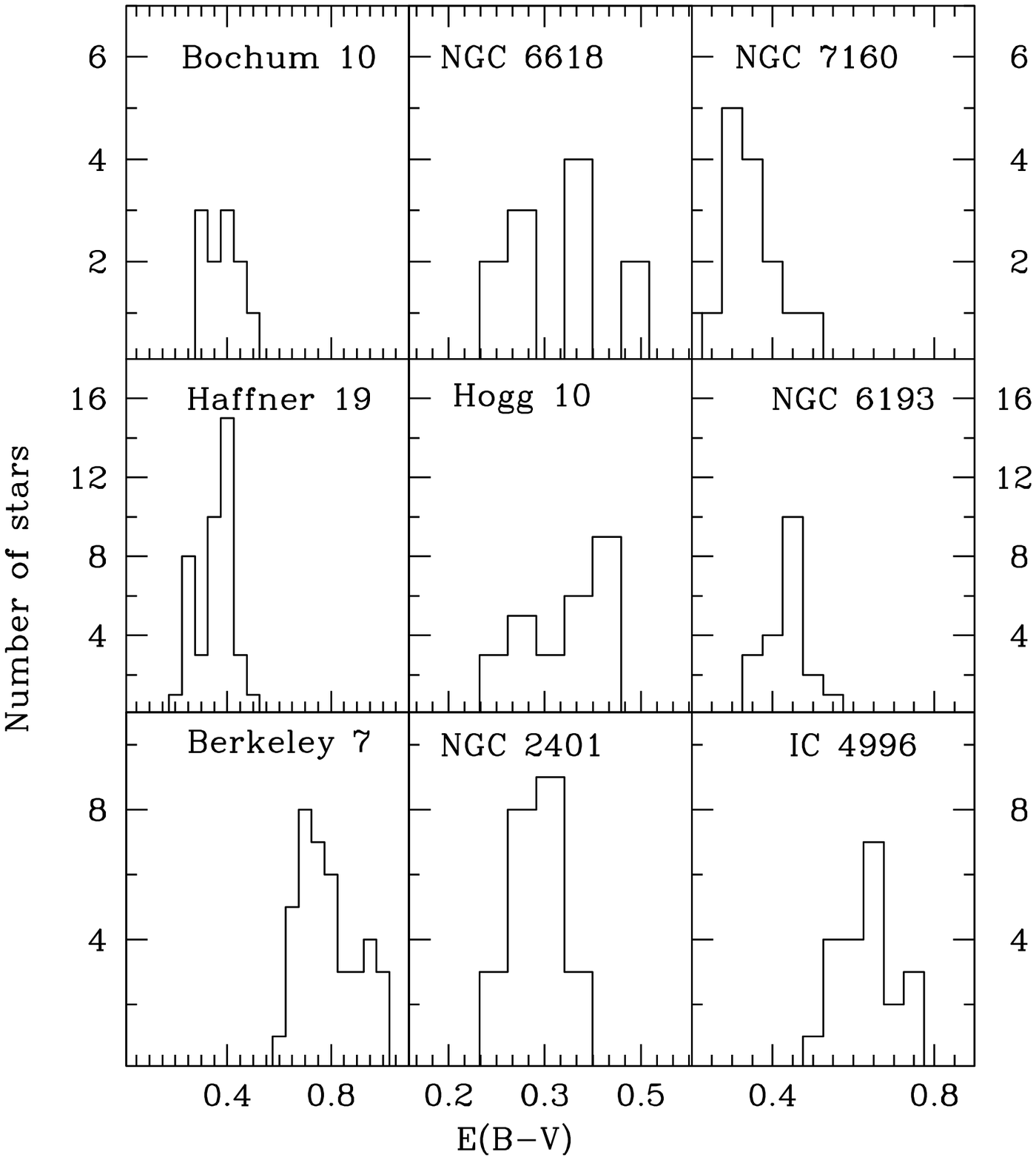}
    \includegraphics[width=9cm,height=7cm]{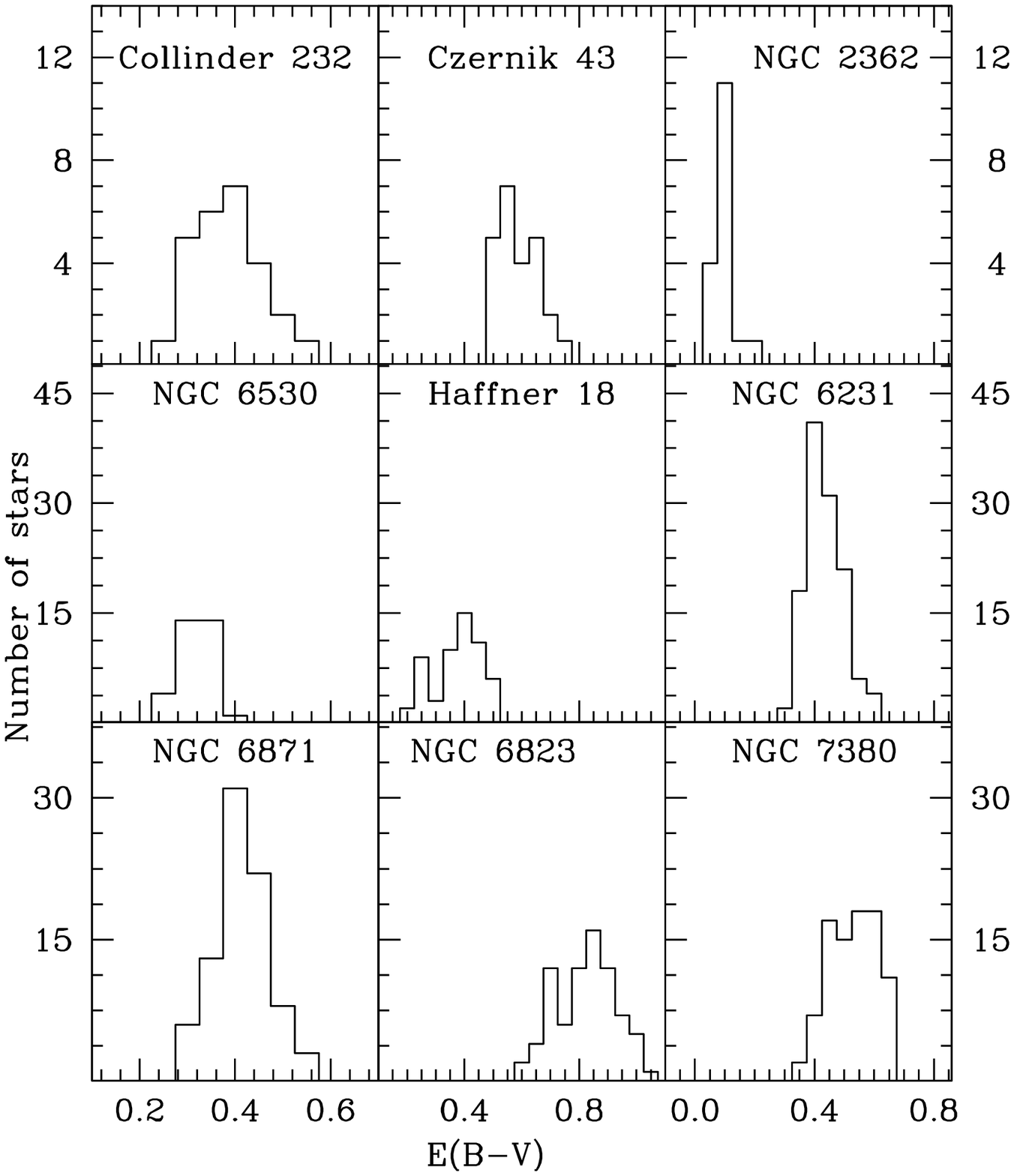}
    \includegraphics[width=6cm,height=4cm]{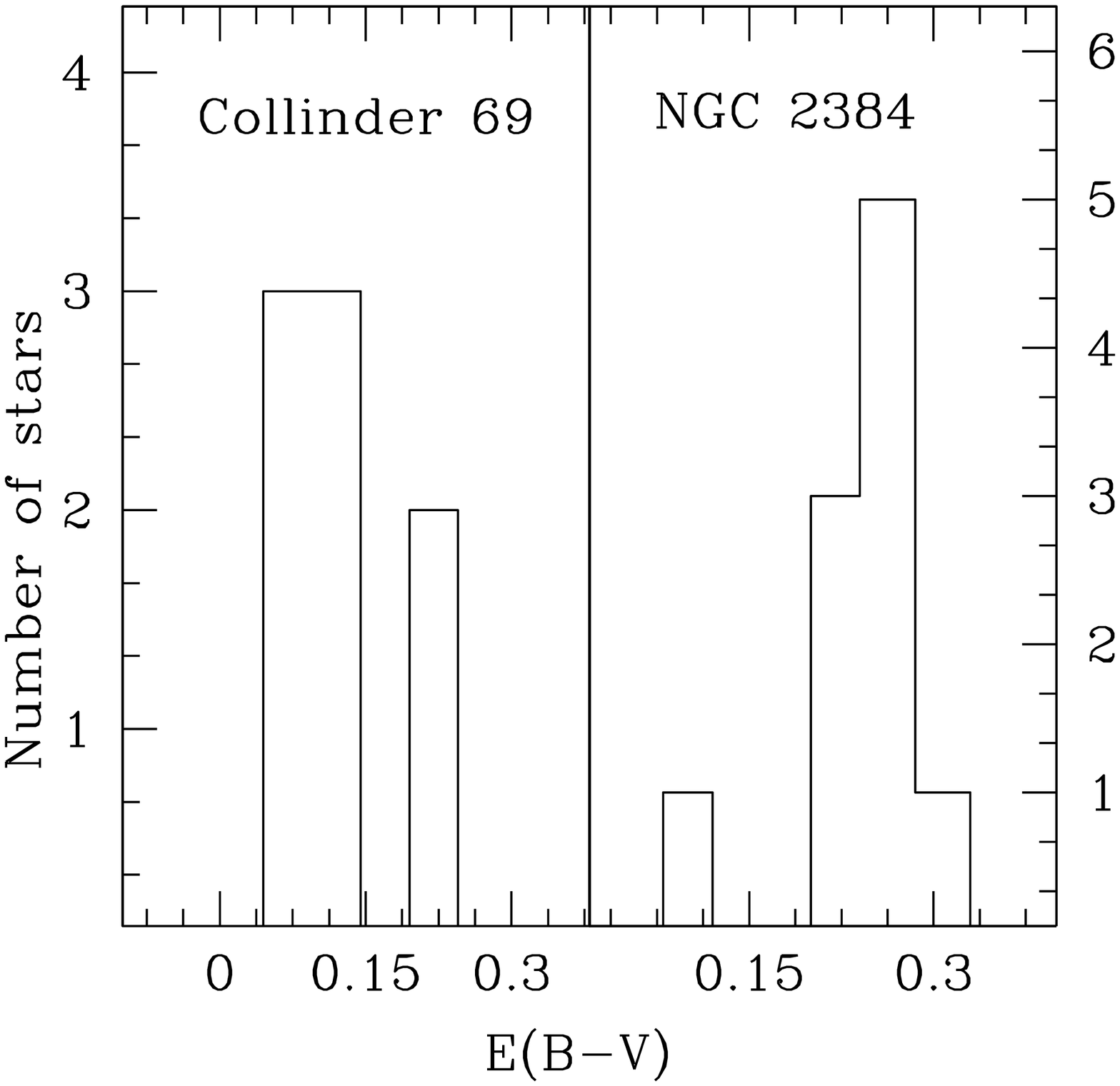}
    \caption{The dispersion of $E(B-V)$ shows the presence of non-uniform extinction
for the clusters under study.}
    \label{hst}
  \end{figure}

\section{Analysis of observational data} \label{sec:ana}

For the study of interstellar extinction law, we have derived the $E(U-V), E(B-V),
E(V-R), E(V-I), E(V-J), E(V-H)$ and $E(V-K)$ colour excesses for early type stars.
We have calculated these excesses by comparing the observed colours with their
intrinsic ones. The intrinsic colours are derived from the MKK spectral type luminosity
class colour relation given by FitzGerald (1970) for $(U-V)$ and $(B-V)$, by Johnson (1966) for $(V-
R)$ and $(V-I)$ and by Koornneef (1983) for $(V-J)$, $(V-H)$ and $(V-K)$. The number of stars
included in the study for a particular cluster are listed in the Table \ref{inf}.

\begin{table*}
\tiny
\centering
\caption{The values of mean $E(B-V)$, $E(B-V)_{min}$, $E(B-V)_{max}$ and $\Delta
E(B-V)$ of each cluster are listed with their central coordinates}
\begin{tabular}{lccccc}
\hline\hline
Cluster & Central Coordinates  & $\overline{E(B-V)}$ & $E(B-V)_{min}$ & $E(B-V)_{max}$ &  $\Delta E(B-V)$      \\
        & $\alpha_{2000}$~~~~$\delta_{2000}$ &     (mag)      &      (mag)  &  (mag)    &   (mag)      \\
\hline
 Berkeley 7   & 01:54:12~$+$62:22:00  &  0.81 & 0.66 &  1.02   &   0.36      \\
 Collinder 69 & 05:35:06~$+$09:56:00  &  0.15 & 0.10 &  0.21   &   0.11      \\
 Hogg 10      & 11:10:42~$-$60:24:00  &  0.40 & 0.30 &  0.49   &   0.19      \\
 NGC 2362     & 07:18:41~$-$24:57:18  &  0.11 & 0.08 &  0.17   &   0.09      \\
 Czernik 43   & 23:25:48~$+$61:19:11  &  0.61 & 0.52 &  0.73   &   0.21      \\
 NGC 6530     & 18:04:31~$-$24:21:30  &  0.34 & 0.27 &  0.40   &   0.13      \\
 NGC 6871     & 20:05:59~$+$35:46:36  &  0.44 & 0.32 &  0.56   &   0.24      \\
 Bochum 10    & 10:42:12~$-$59:08:00  &  0.41 & 0.34 &  0.48   &   0.14      \\
 Haffner 18   & 07:52:39~$-$26:23:00  &  0.40 & 0.24 &  0.53   &   0.29      \\
 IC 4996      & 20:16:30~$+$37:38:00  &  0.67 & 0.56 &  0.77   &   0.21      \\
 NGC 2384     & 07:25:10~$-$21:01:18  &  0.27 & 0.22 &  0.31   &   0.09      \\
 NGC 6193      & 16:41:20~$-$48:45:48  &  0.47 & 0.39 &  0.53   &   0.14      \\
 NGC 6618     & 18:20:47~$-$16:10:18  &  0.39 & 0.31 &  0.48   &   0.17      \\
 NGC 7160     & 21:53:40~$+$62:36:12  &  0.39 & 0.32 &  0.48   &   0.16      \\
 Collinder 232     & 10:44:59~$-$59:33:00  &  0.41 & 0.30 &  0.52   &   0.22      \\
 Haffner 19   & 07:52:47~$-$26:17:00  &  0.38 & 0.25 &  0.48   &   0.23      \\
 NGC 2401     & 07:29:24~$-$13:58:00  &  0.35 & 0.29 &  0.41   &   0.12      \\
 NGC 6231     & 16:54:10~$-$41:49:30  &  0.46 & 0.35 &  0.63   &   0.28      \\
 NGC 6823     & 19:43:09~$+$23:18:00  &  0.86 & 0.66 &  1.04   &   0.38      \\
 NGC 7380     & 22:47:21~$+$58:07:54  &  0.56 & 0.40 &  0.69   &   0.29      \\
\hline
\end{tabular}
\label{del}
\end{table*}

\subsection {Non-uniform extinction}

To study the presence of non-uniform extinction in each cluster we have calculated
$\Delta E(B-V) = E(B-V)_{max} - E(B-V)_{min}$, where $E(B-V)_{max}$ and $E(B-V)_{min}$
are the average $E(B-V)$ of four highest and four lowest values of $E(B-V)$.
The values of $\Delta E(B-V)$ along with the central coordinates of each clusters are
listed in the Table \ref{del}. The factors which can produce dispersion in $\Delta E(B-V)$
other than non-uniform extinction can cause maximum dispersion
$\sim$ 0.11 mag (Burki 1975, Sagar 1985, Sagar 1987, Pandey et al. 1990 and Yadav \&
Sagar 2001). Therefore, the value of $\Delta E(B-V)$  greater than
0.11 mag can be considered as non-uniform extinction in the cluster. In the present study
except three clusters namely NGC 2362, NGC 2384 and Collinder 69,  all are having
$\Delta E(B-V)$ value more than 0.11 mag. So this analysis confirms the presence of
non-uniform extinction in all clusters except NGC 2362, NGC 2384 and Collinder 69.
In case of NGC 6530 we found a result similar as that reported by Sagar (1987).

In addition to this, we have also analyzed the extent of non-uniform extinction by
plotting the histograms of $E(B-V)$ as shown in Fig. \ref{hst} for each cluster.
Histograms show that all the clusters have a wide range in $E(B-V)$ values. This 
indicates that all the clusters have different amounts of non-uniform extinction and 
dust is non-uniformly distributed over the cluster region. In the following sections
we explore the possible reasons for the presence of this type of distribution in $E(B-V)$.

\subsection{Extinction law} \label{sec:color}

\begin{table*}
\tiny
\centering
\caption{The derived and normal colour excess ratios for all the clusters under study. }
\begin{tabular}{llllllll}
\hline\hline
Cluster & $E(U-V)$ & $E(V-R)$ & $E(V-I)$ & $E(V-J)$ & $E(V-H)$ & $E(V-K)$    \\
        & $\overline{E(B-V)}$ & $\overline{E(B-V)}$  &$\overline{E(B-V)}$  & $\overline{E(B-V)}$  & $\overline{E(B-V)}$  & $\overline{E(B-V)}$    \\
\hline
 Normal value  &  1.72       &   0.60    &     1.25    &    2.30     &     2.58    &   2.78          \\
 &&&&&&\\
 Berkeley 7 &   1.72$\pm$0.03 &   0.67$\pm$0.02  &  1.18$\pm$0.05  &  1.98$\pm$0.13 &   2.48$\pm$0.13  &  2.61$\pm$0.13   \\
NGC 6823   &   1.74$\pm$0.03 &   0.67$\pm$0.05  &  1.39$\pm$0.16  &  2.10$\pm$0.23 &   2.44$\pm$0.18  &  2.64$\pm$0.22    \\
 Collinder 232 & 1.88$\pm$0.06 &   0.62$\pm$0.05  &  1.37$\pm$0.13  &  2.55$\pm$0.37 &   2.98$\pm$0.52 &  3.18$\pm$0.57    \\
 NGC 6530   &   1.87$\pm$0.17 &   0.81$\pm$0.15  &  1.57$\pm$0.16  &  3.21$\pm$0.13 &   3.55$\pm$0.46  &  4.19$\pm$0.48    \\
 Haffner 19 &   1.73$\pm$0.09 &   0.61$\pm$0.04  &  1.42$\pm$0.10  &  2.10$\pm$0.25 &   2.91$\pm$0.17  &  3.10$\pm$0.28    \\
 NGC 7160   &   2.00$\pm$0.19 &   0.64$\pm$0.20  &  1.35$\pm$0.23  &  2.14$\pm$0.29 &   2.09$\pm$0.35  &  2.50$\pm$0.27    \\
 NGC 6193   &   1.86$\pm$0.18 &   0.68$\pm$0.07  &  1.59$\pm$0.18  &  2.03$\pm$0.37 &   2.68$\pm$0.44  &  2.86$\pm$0.54    \\
 Bochum 10  &   2.38$\pm$0.30 &   0.64$\pm$0.15  &  1.40$\pm$0.55  &  2.17$\pm$0.64 &   2.37$\pm$0.60  &  3.20$\pm$0.69    \\
 Haffner 18 &   1.76$\pm$0.03 &   0.61$\pm$0.03  &  1.30$\pm$0.09  &  2.06$\pm$0.23 &   2.78$\pm$0.13  &  2.93$\pm$0.23    \\
 NGC 2362   &   2.25$\pm$0.23 &   0.47$\pm$0.13  &  1.15$\pm$0.28  &  2.41$\pm$0.86 &   2.68$\pm$0.78  &  3.94$\pm$0.74    \\
 Hogg 10    &   1.99$\pm$0.06 &   0.60$\pm$0.08  &  1.39$\pm$0.09  &  2.07$\pm$0.25 &   2.27$\pm$0.38  &  2.66$\pm$0.45    \\
 NGC 2401   &   1.84$\pm$0.14 &   0.77$\pm$0.11  &  1.53$\pm$0.18  &  2.62$\pm$0.36 &   2.88$\pm$0.28  &  2.63$\pm$0.32    \\
 IC 4996    &   1.75$\pm$0.10 &   0.33$\pm$0.06  &  1.86$\pm$0.12  &  1.70$\pm$0.32 &   2.18$\pm$0.42  &  2.64$\pm$0.45    \\
 NGC 2384   &   1.65$\pm$0.21 &   0.33$\pm$0.28  &  2.34$\pm$0.87  &  2.33$\pm$0.58 &   2.29$\pm$0.39  &  2.76$\pm$0.33    \\
 Czernik 43 &   1.78$\pm$0.05 &   0.54$\pm$0.10  &       *         &  1.55$\pm$0.23 &   2.10$\pm$0.24  &  2.22$\pm$0.26    \\
 NGC 6871   &   1.68$\pm$0.05 &        *         &       *         &  2.02$\pm$0.27 &   2.37$\pm$0.19  &  2.50$\pm$0.20    \\
NGC 6618   &   1.91$\pm$0.06 &        *         &       *         &  1.93$\pm$0.21 &   2.40$\pm$0.52  &  2.83$\pm$0.82    \\
 NGC 7380   &   1.76$\pm$0.03 &        *         &       *         &  2.11$\pm$0.16 &   2.68$\pm$0.15  &  2.78$\pm$0.19    \\
 NGC 6231   &   1.85$\pm$0.04 &        *         &  1.24$\pm$0.03  &  2.15$\pm$0.16 &   2.64$\pm$0.14  &  2.75$\pm$0.16    \\
 Collinder 69 & 1.79$\pm$0.10 &        *         &       *         &  2.45$\pm$0.57 &   2.30$\pm$0.36  &  3.86$\pm$0.36          \\
\hline
\end{tabular}
\label{ex}
\end{table*}

\begin{figure}
    \centering
    \includegraphics[width=8cm,height=8cm]{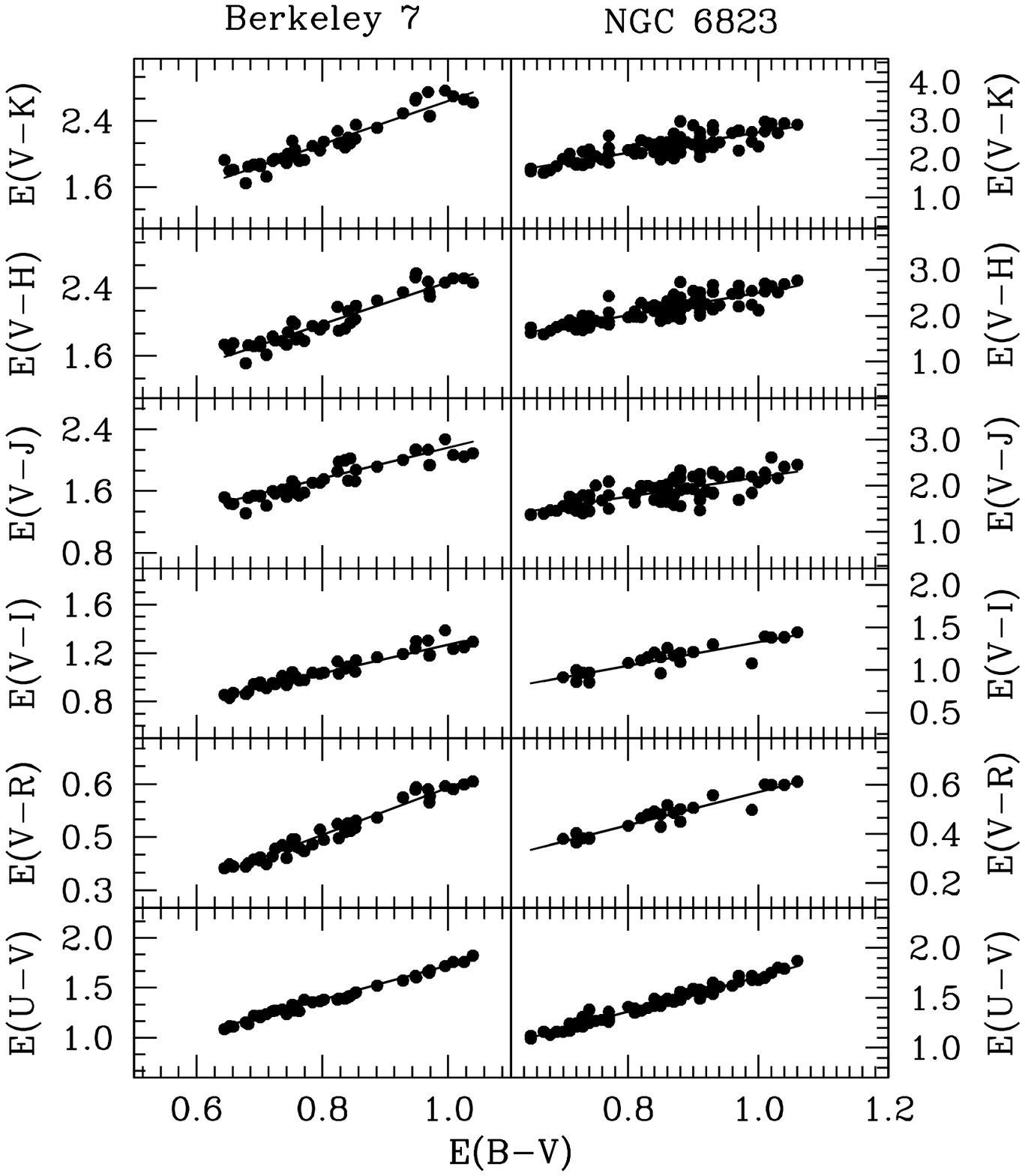}
    \includegraphics[width=8cm,height=8cm]{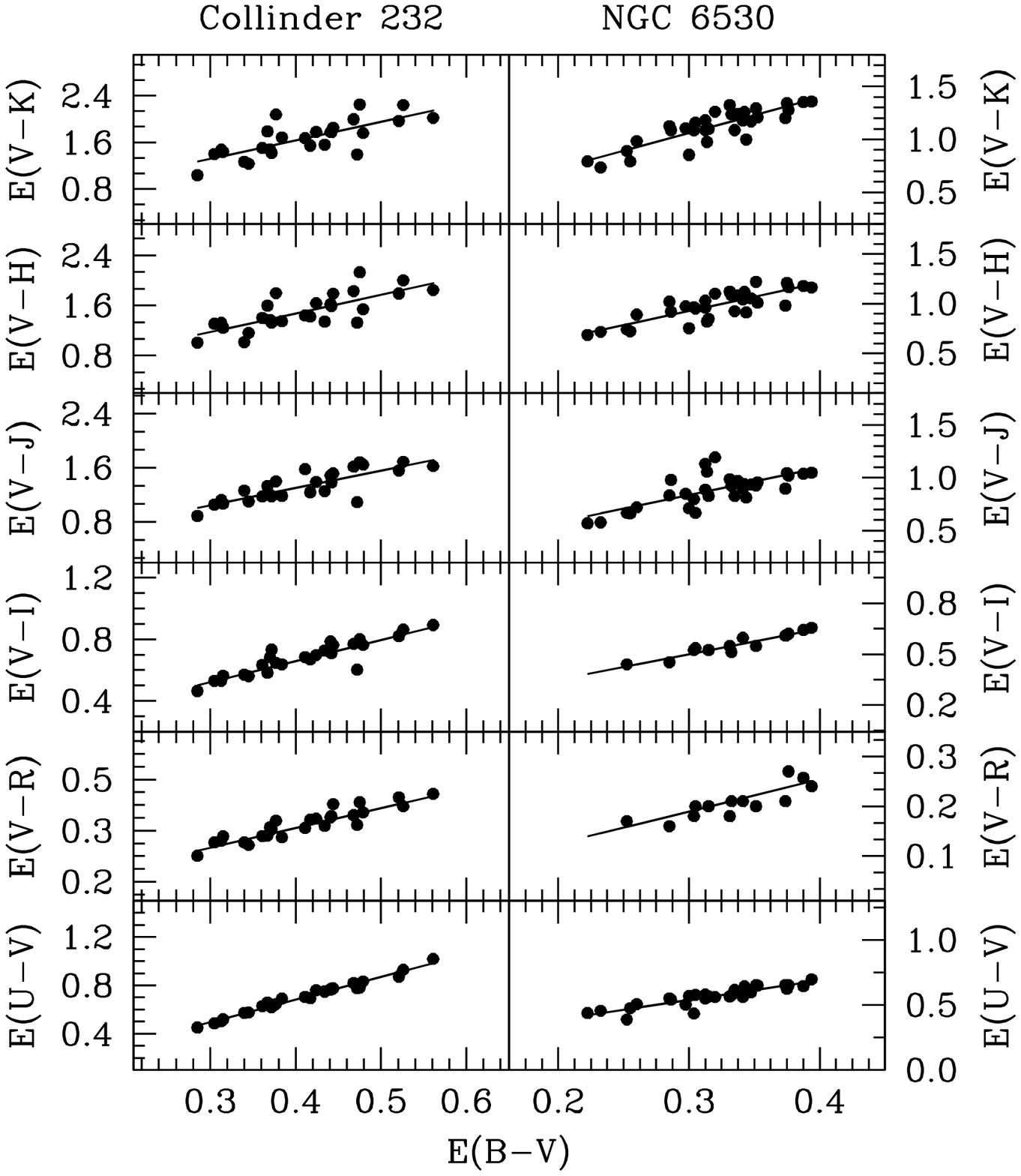}
    \caption{The colour excesses $E(U-V), E(V-R), E(V-I), E(V-J), E(V-H), E(V-K)$
    are plotted against $E(B-V)$ for the clusters Berkeley 7, NGC 6823, Collinder
    232 and NGC 6530.}
    \label{excess1}
  \end{figure}

\begin{figure}
    \centering
    \includegraphics[width=8cm,height=8cm]{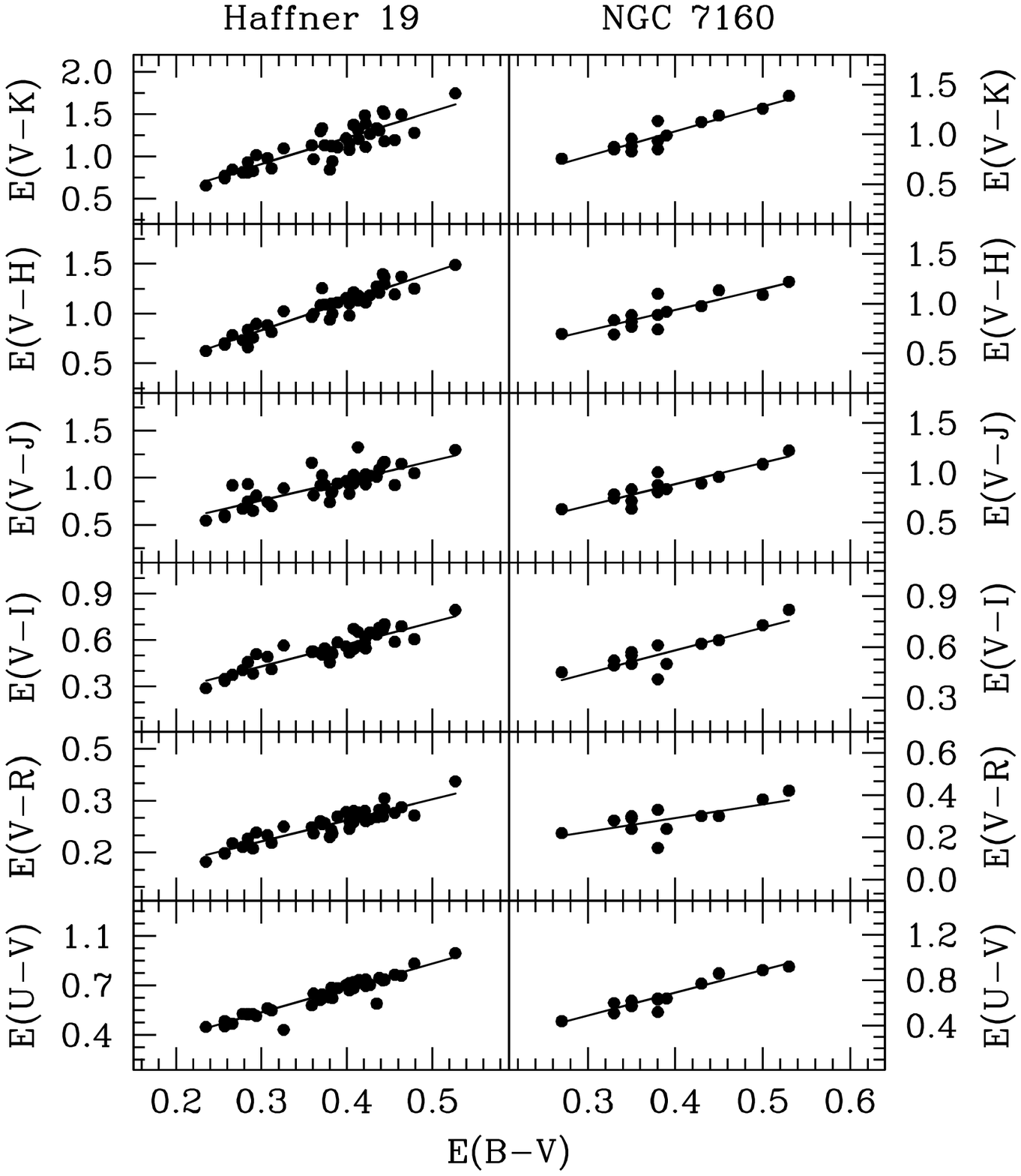}
    \includegraphics[width=8cm,height=8cm]{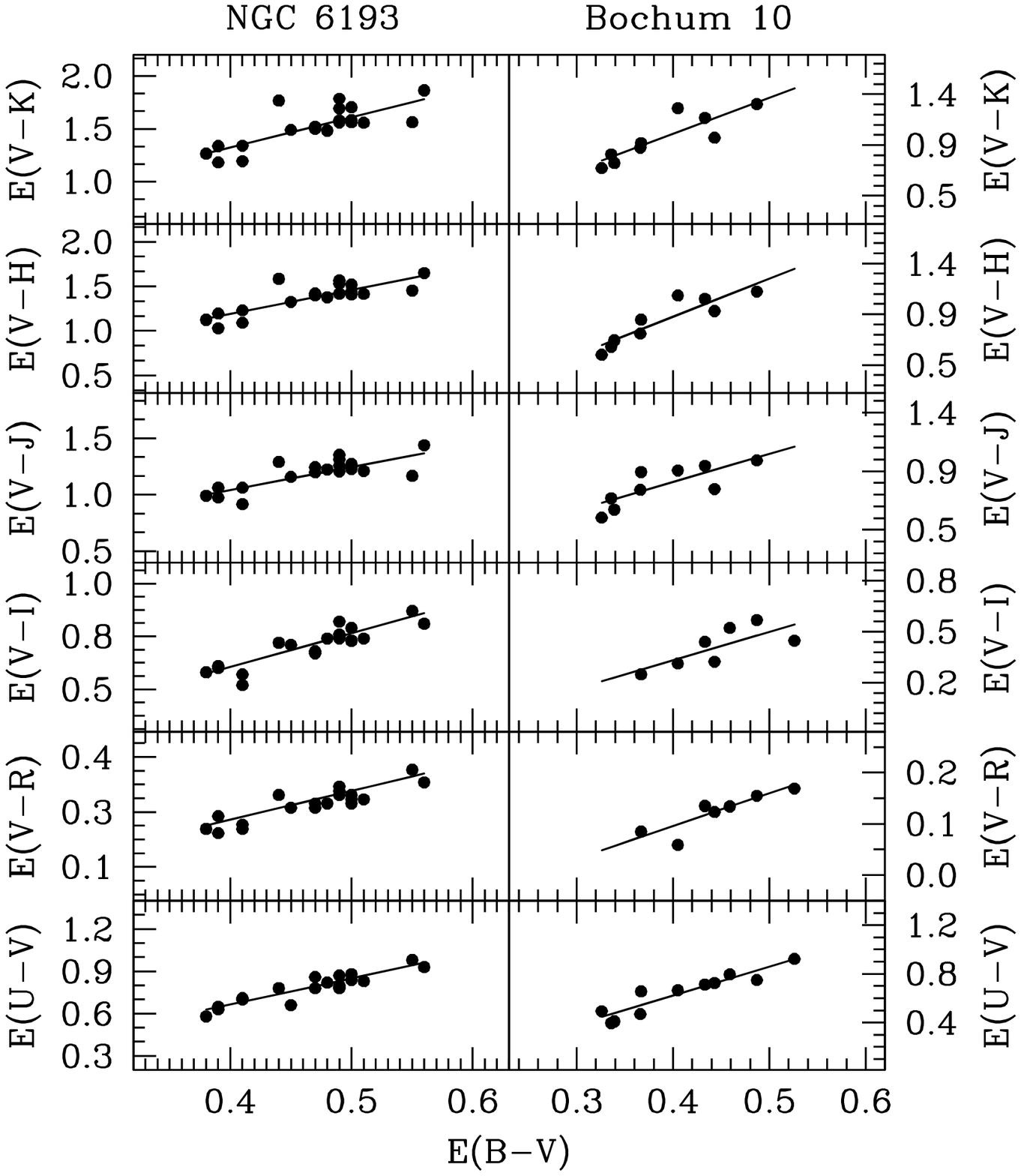}
    \caption{Same as Fig. \ref{excess1} for the clusters Haffner 19, NGC 7160, NGC 6193
    and Bochum 10.}
    \label{excess2}
  \end{figure}

\begin{figure}
    \centering
    \includegraphics[width=8cm,height=8cm]{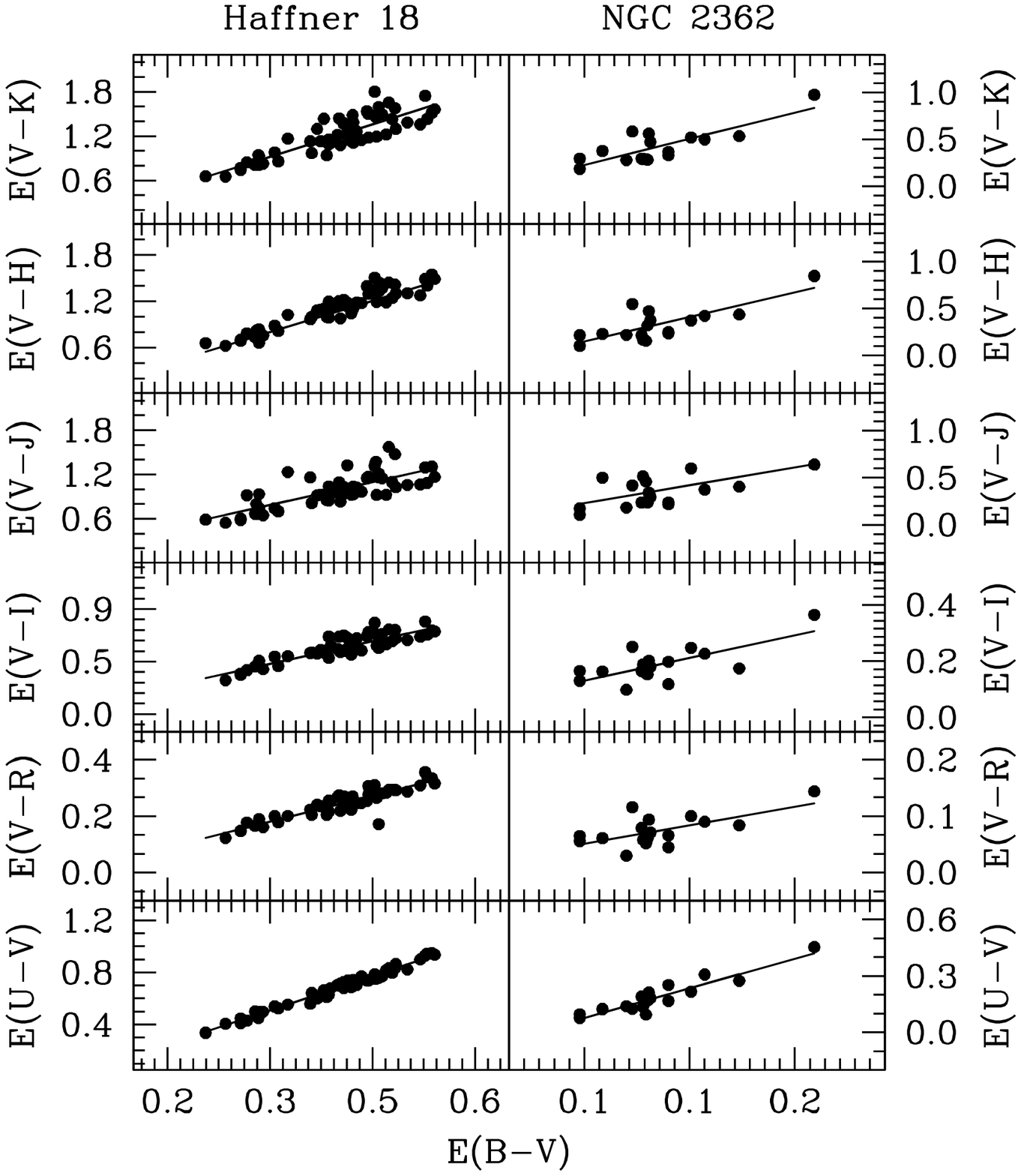}
    \includegraphics[width=8cm,height=8cm]{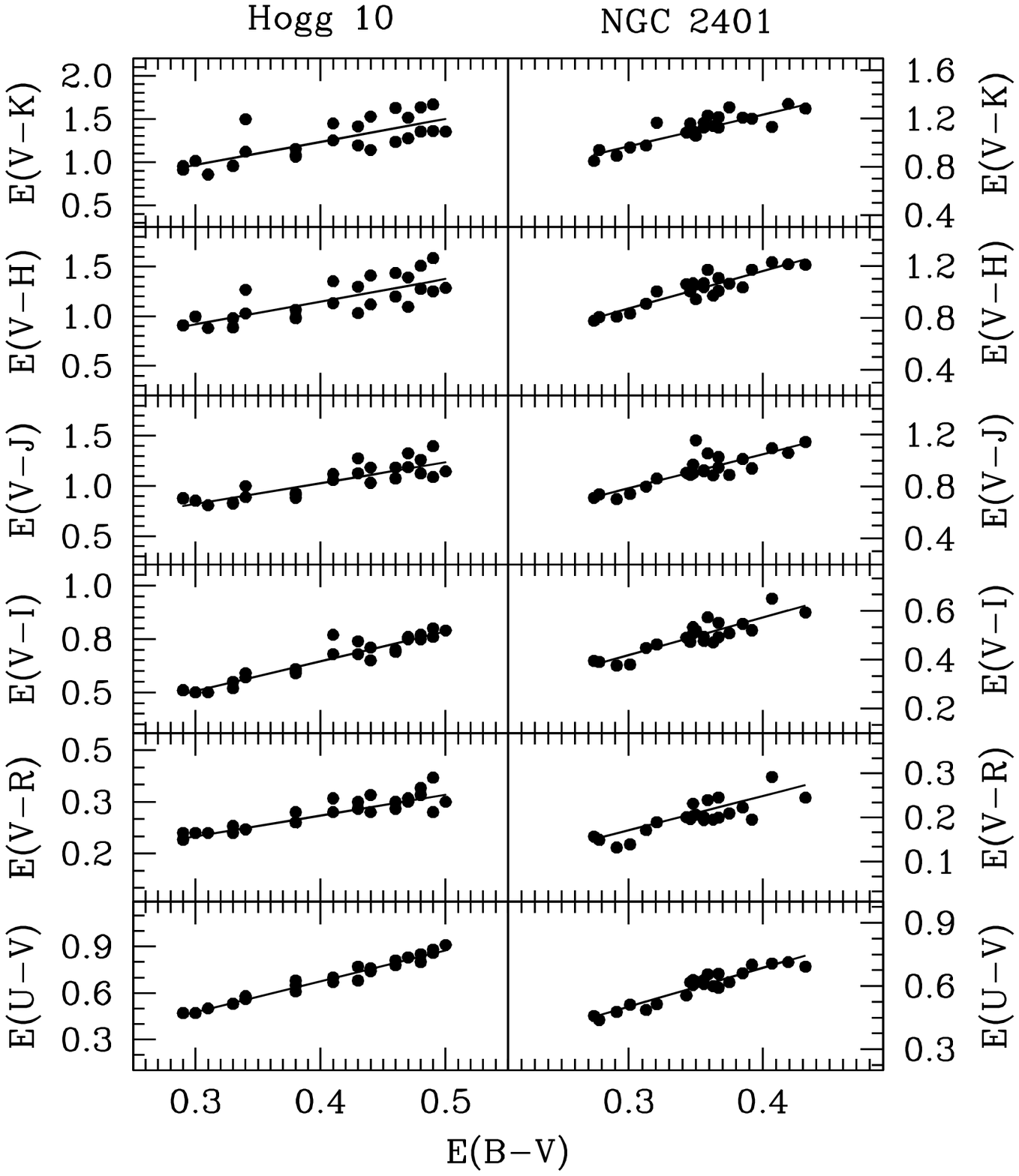}
    \caption{Same as Fig. \ref{excess1} for the clusters Haffner 18, NGC 2362, Hogg 10 and
    NGC 2401.}
    \label{excess3}
  \end{figure}

\begin{figure}
    \centering
    \includegraphics[width=8cm,height=8cm]{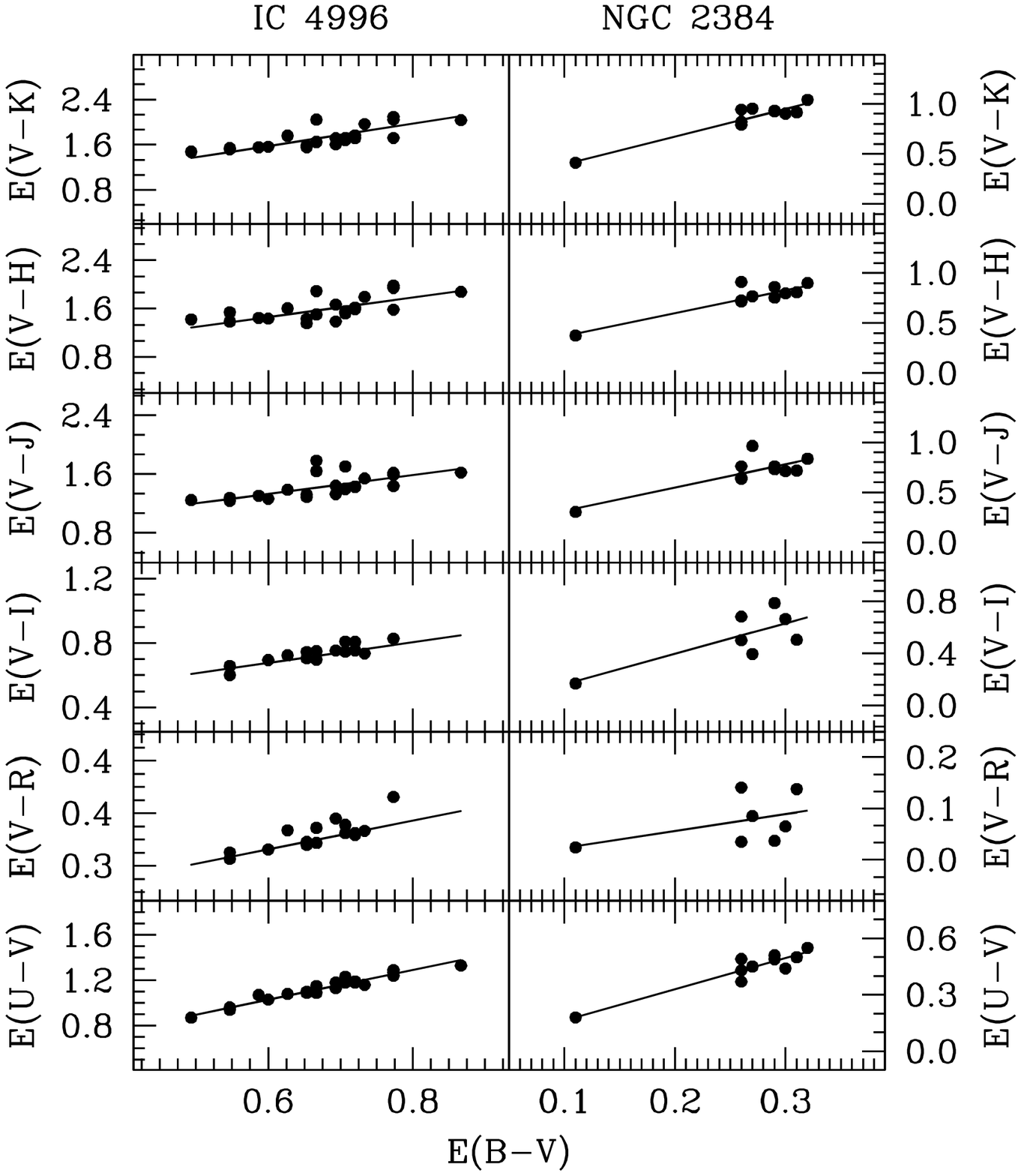}
    \includegraphics[width=8cm,height=8cm]{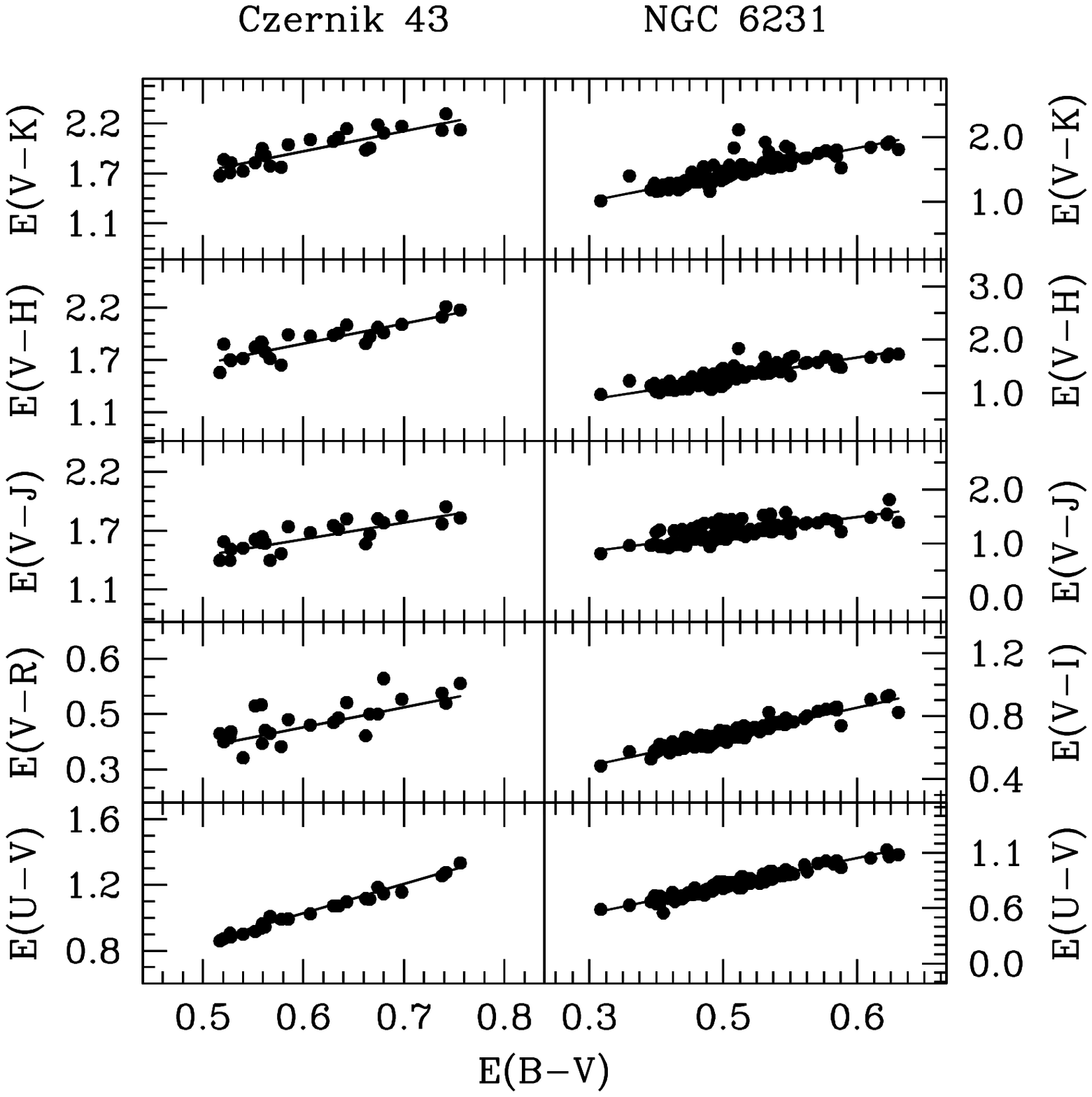}
    \caption{Same as Fig. \ref{excess1} for the clusters IC 4996, NGC 2384 Czernik 43 and NGC 6231.}
    \label{excess4}
  \end{figure}

\begin{figure}
    \centering
    \includegraphics[width=8cm,height=9cm]{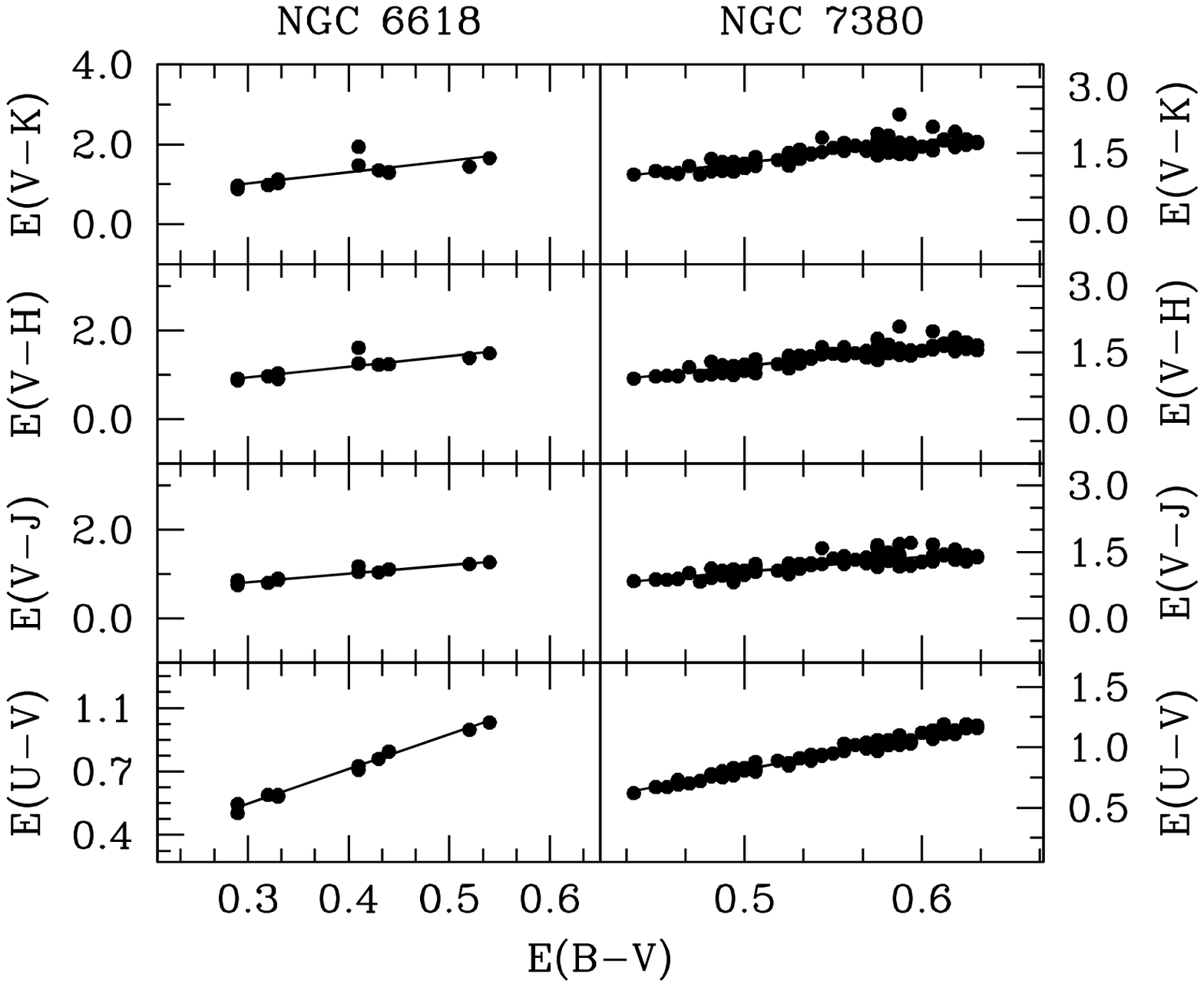}
    \includegraphics[width=8cm,height=9cm]{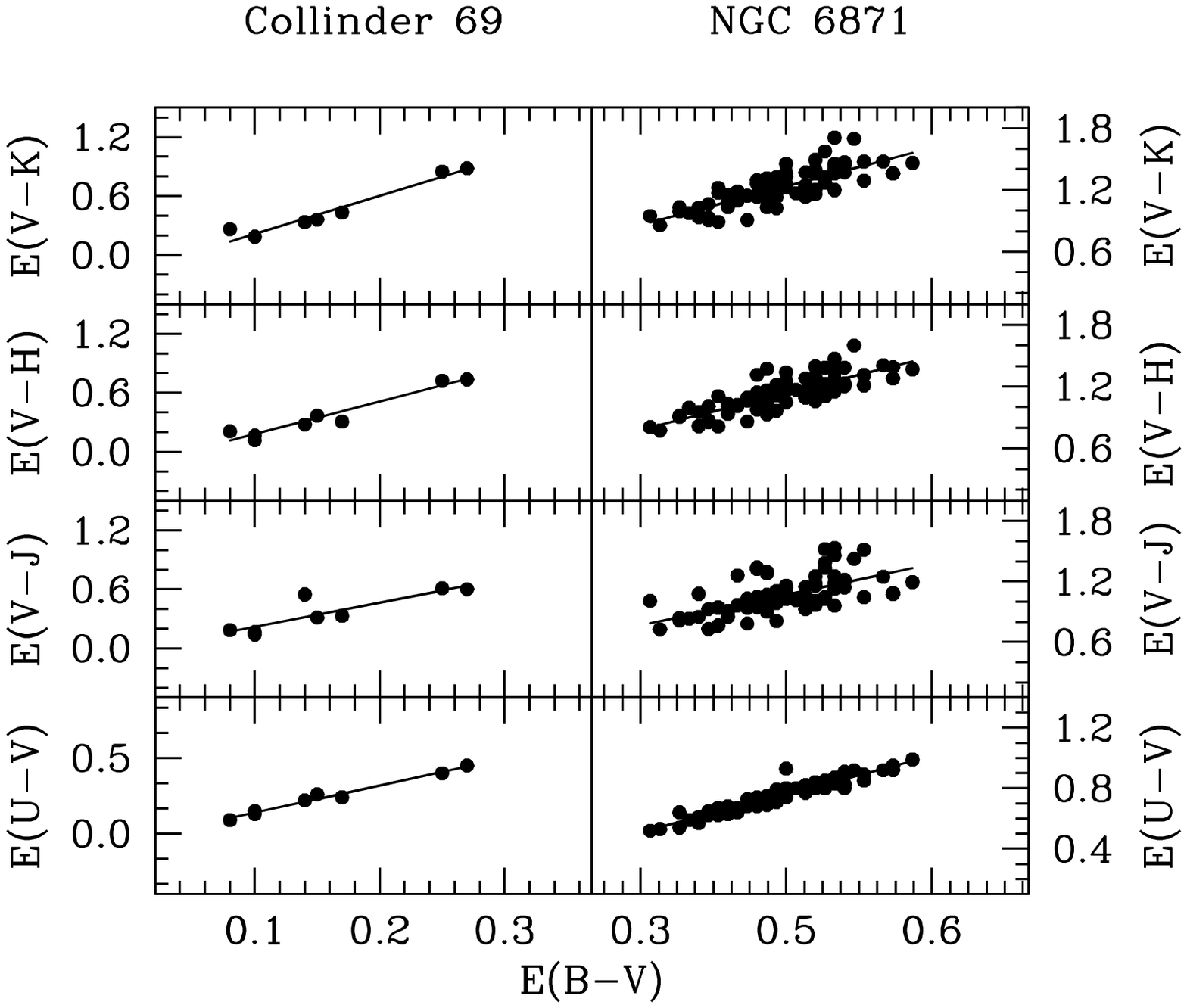}
    \caption{Same as Fig. 2 for the clusters NGC 6618, NGC 7380, Collinder 69 and NGC 6871.}
    \label{excess5}
  \end{figure}

To investigate the interstellar extinction law towards the clusters, we have plotted the colour
excesses $E(U-V)$, $E(V-R)$, $E(V-I)$, $E(V-J)$, $E(V-H)$ and $E(V-K)$ against $E(B-V)$ in Fig
\ref{excess1}, \ref{excess2}, \ref{excess3}, \ref{excess4} and \ref{excess5}. The solid line in
the plots is the least-square fit to the data points. The slope of the colour excess ratios alongwith
their error are listed in Table \ref{ex}. The theoretical values of the colour excess ratios are
taken from Pandey et al. (2002). By comparing the observed colour excess ratios with the theoretical
one, following  conclusions can be drawn.

(i) Twelve clusters namely NGC 6823, Haffner 18, Haffner 19, NGC 7160, NGC 6193, NGC 2401,
NGC 2384, NGC 6871, NGC 7380, Berkeley 7, Collinder 69 and IC 4996 follow the normal extinction
law in optical and near-IR region within 2$\sigma$, where $\sigma$ is the standard deviation of
data points. This shows the expected behaviour of the normal interstellar law in the direction of these clusters
and hence presumably dust with normal characteristics in shape, size and composition (Tapia et al. 1991).
Similar results have also
been discussed in Yadav \& Sagar (2001) for the clusters NGC 1893, NGC 2244, Tr 37 and Be 68.

(ii) Six clusters namely Bochum 10, NGC 2362, Collinder 232, Hogg 10, NGC 6618 and NGC 6231 show
anomalous behaviour in $\lambda < \lambda_{J}$. These clusters have colour excess ratios greater
than that of the normal value in optical band. This indicates that extinction law is different
than for the normal extinction law. This also implies that dust particles in the direction of these clusters have
smaller cross section than the normal ones (Turner 1994).

(iii) Two clusters NGC 6530 and Czernik 43 show anomalous behaviour for $\lambda \geq \lambda_{J}$.
NGC 6530 have large colour excess ratios while Czernik 43 have less ratios than the normal one.

The extinction law around early type stars can be understood by the model given by Seab \&
Shull's (1983) for dust grain processing due to passage of shock waves. Results of this model
are naturally dependent on the physical conditions of the intracluster material (Sagar \&
Qian 1993). Small inhomogeneities in them would therefore result in large differences in the
observed parameters. Consequently, depending upon the physical conditions of intracluster 
material, grain size distribution resulting from the interactions of strong radiations of
hot O and B stars can be either normal or shifted by a small amount in either direction
compared to that of normal sized particles. The anomaly depends on the direction of the
shift, which ultimately depends on the grain destruction mechanism.

\subsection{Variation of colour excess with age of the clusters} \label{sec:age}

\begin{figure}
    \centering
    \includegraphics[width=9cm,height=8cm]{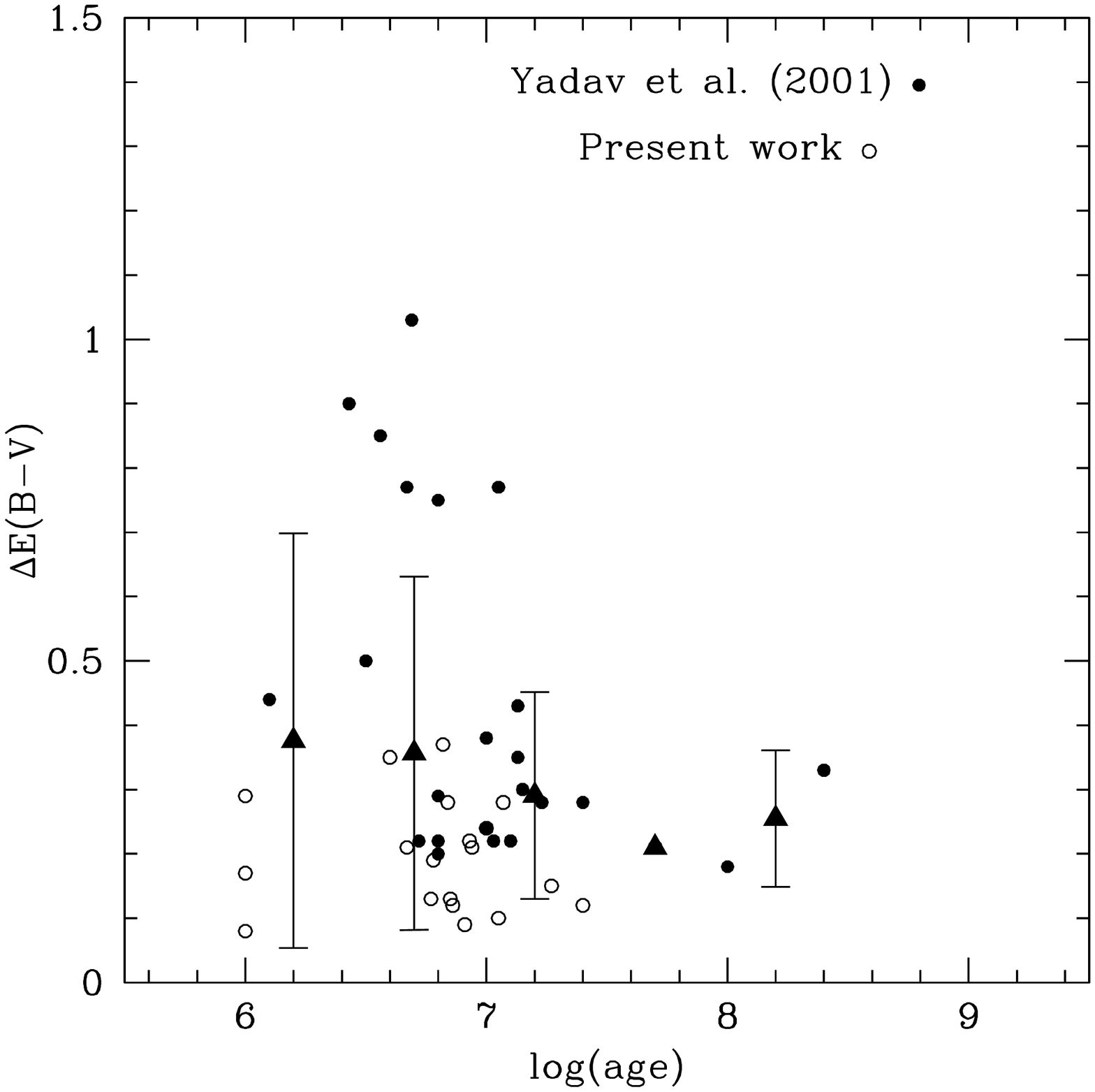}
    \caption{Variation of $\Delta E(B-V)$ with the cluster age. The open circles
    denotes data points from present work and the filled circles denotes data points from Yadav
    et al. (2001)}. Error bars denote the corresponding standard deviation of binned data.
    \label{age2}
  \end{figure}

It is believed that stars are formed in the dense clouds of dust and gas, by contracting
under their own gravitational attraction (Yorke \& Krugel, 1977 and Pandey et al., 1990).
Not all of this gas or dust becomes
part of star but also used in the formation of planets, asteroids, comets etc. or
remains as dust. So this remnant dust is either used in the star formation or
blown away by the high radiation pressure of embedded or nearby stars (Pandey et al., 1990).
Therefore, young open star clusters should have some amount of dust and gas
which gradually decreases with time. It means there should be a
relationship between age of the star cluster and the interstellar extinction.
Pandey et al. (1990) studied the variation of non-uniform extinction $\Delta E(B-V)$ with
age for 64 open star clusters and found that extinction decreases with the age of the clusters
and calculated the gas removal time larger than $10^{8}$ yrs.
As $\Delta E(B-V)$ represents the measure of non-uniform extinction, so we have plotted the
age of each cluster with $\Delta E(B-V)$ to see the relationship between age of cluster
and the presence of non-uniform extinction. This plot is shown in Fig. \ref{age2}. The values of
$\Delta E(B-V)$ of the clusters under study are listed in the Table \ref{del} and the ages of
the clusters are taken from Table \ref{inf}. For this study, we have also included the clusters studied by
Yadav \& Sagar (2001). To see the variation of mean $\Delta E(B-V)$ with age, we have binned the data
into several groups with an age interval of 0.5 Myr. After binning the data we have
calculated the mean of $\Delta E(B-V)$ and its standard deviation, which are plotted in Fig. \ref{age2}.
The variation of $\Delta E(B-V)$ with age indicates that a large amount of
obscuring material is present in younger age clusters. For older clusters the mean value of $\Delta E(B-V)$
is small. So we can conclude that the extent of non-uniform extinction decreases with age.
Dust and gas  may be blown away from the older clusters due to several reasons as discussed in
Yadav and Sagar (2001). A similar analysis was also done by Yadav and Sagar (2001)
and found the similar relationship between $\Delta E(B-V)$ and age of the clusters.

\subsection{Spatial variation of colour excess} \label{sec:spt}

To study the spatial variation of colour excess $E(B-V)$, we have divided the whole
region of the clusters into small square boxes. The clusters Collinder 69,
Collinder 232, Czernik 43, NGC 6530, Bochum 10, NGC 2384, NGC 6618 and NGC 7160 were not
considered for this study due to insufficient number of stars and smaller area. The area of the
clusters Berkeley 7, Haffner 18, Haffner 19, Hogg 10, IC 4996, NGC 2401, NGC 6193,
NGC 6231 and NGC 6823 is divided into small boxes of area $5^{\prime} \times 5^{\prime}$.
For the clusters NGC 6871 and NGC 7380 the square boxes are of $10^{\prime} \times 10^{\prime}$
due to the available larger area. We have calculated the mean value of $E(B-V)$
and its standard deviation for the stars in each boxes. The close inspection of these
values provides a clue that there is no significant variation of the mean colour excess
values with the positions in the cluster area. A small variation of mean $E(B-V)$ with
position in NGC 6193 is observed and listed in Table \ref{spatial}. In NGC 6193 the value of
$E(B-V)$ decreases from east to west. A variation in $E(B-V)$ with position has also been
observed by Sagar (1987) for three clusters NGC 6611,
IC 1805 and NGC 6530 and by Yadav \& Sagar (2001) for Tr 14.

\begin{table}
\tiny
\centering
\caption{The spatial variation of $E(B-V)$ across the open cluster NGC 6193. In each
boxes, there is mean values of $E(B-V)$ with their standard deviation for corresponding
$5\times 5$ arcmin$^{2}$. The numbers in the bracket represent the number of stars in that box.}

\begin{tabular}{cccccc}
\hline
$\Delta\alpha$ / $\Delta\delta$   &  -5 - 0   &    0 - 5    &      5 - 10    &      10 - 15    &    15
 - 20     \\
\hline
-10 - -5    &    -            &  0.46(1)        & -                &  -          
&  -     \\
 -5 -  0  & 0.39$\pm$0.01(3) &        -         & 0.49$\pm$0.04(5)&  0.52$\pm$0.04(2)
 &      -     \\
  0 -  5  & 0.40$\pm$0.01(2) &  0.50(1)  & 0.48$\pm$0.01(5)  &      -          & 
  0.50(1)     \\
\hline
\end{tabular}
\label{spatial}
\end{table}

\subsection{Spectral type variation of colour excess} \label{sec:sprl}

\begin{figure}
    \centering
    \includegraphics[width=9cm,height=8cm]{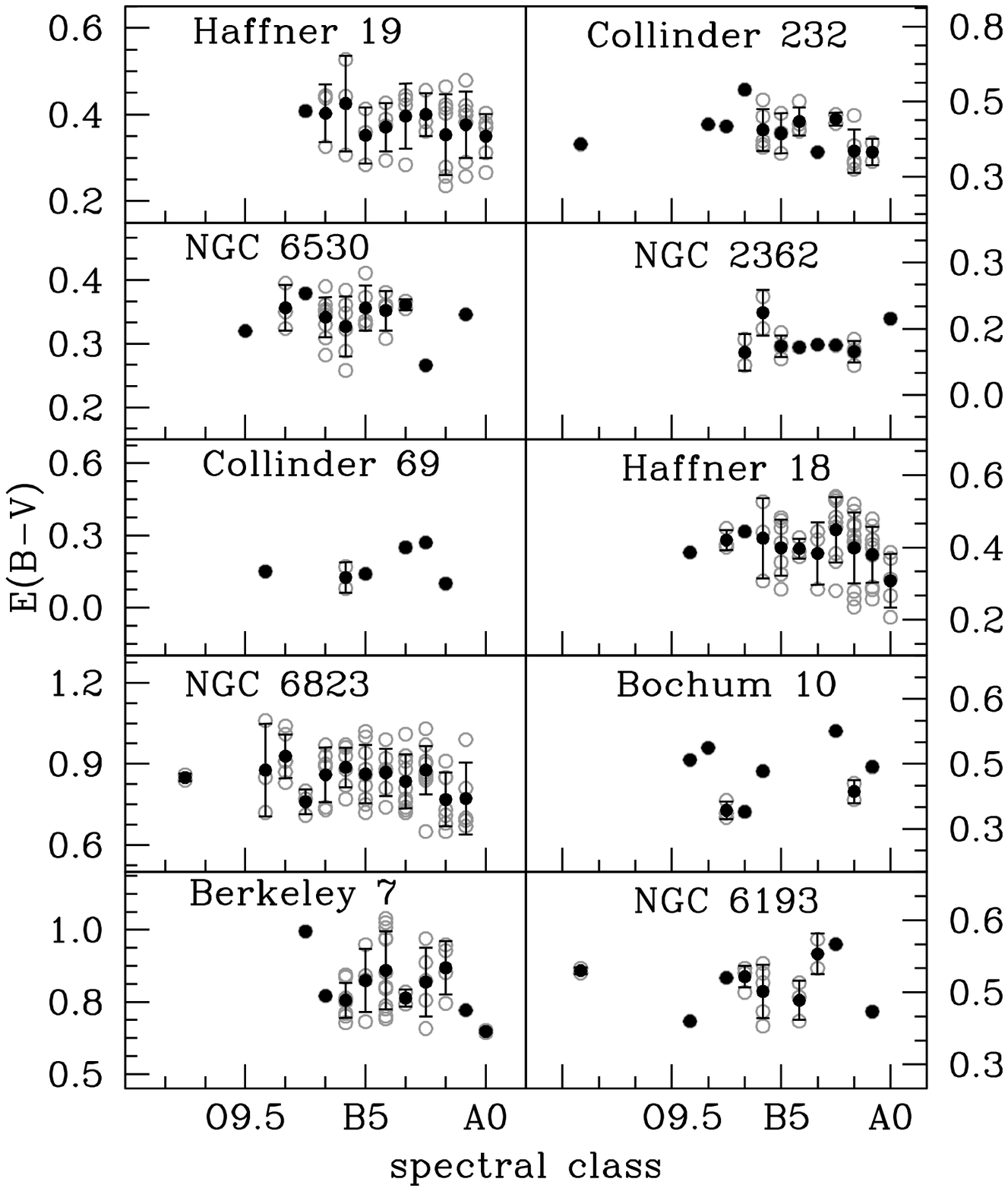}
    \includegraphics[width=9cm,height=8cm]{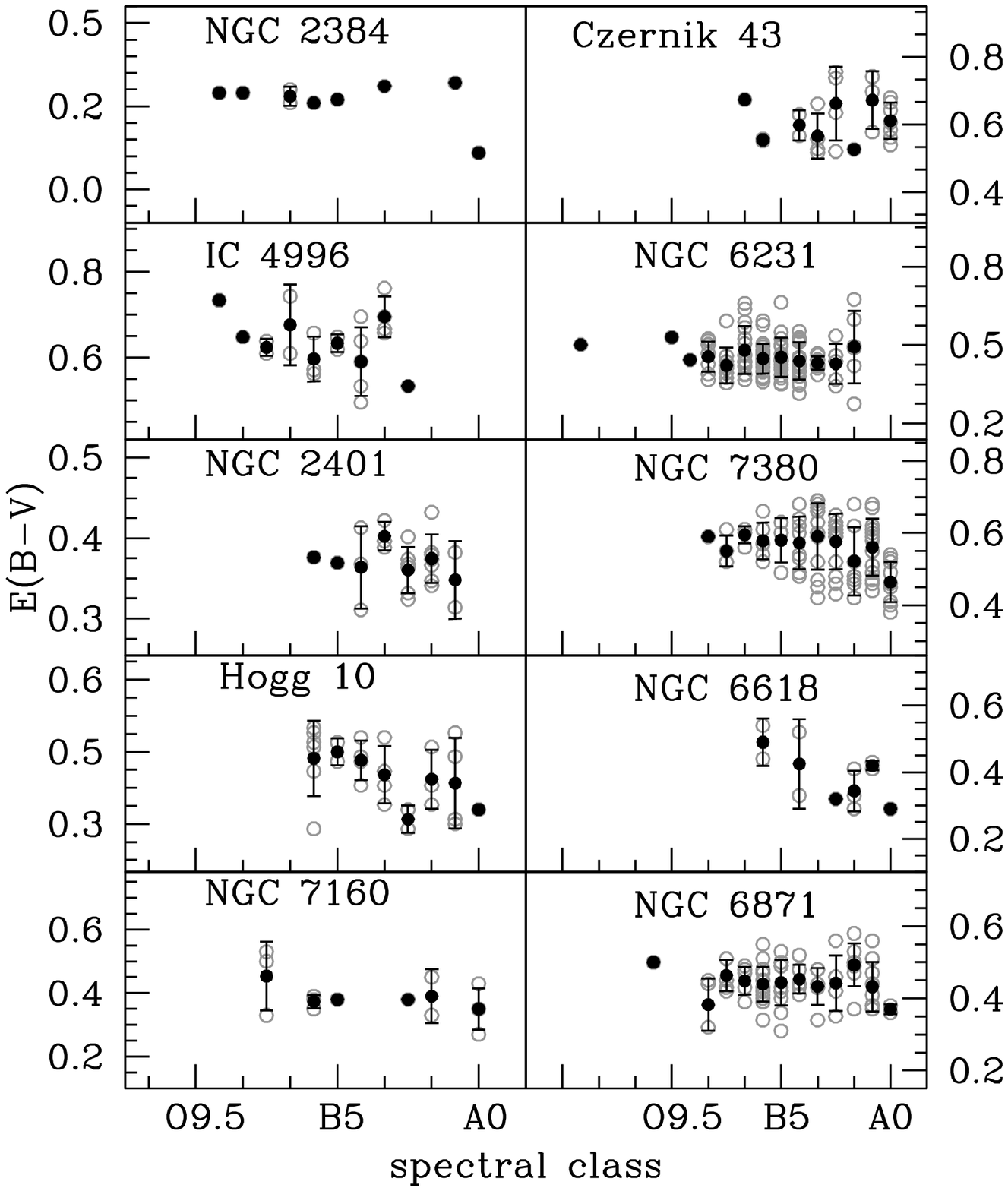}
    \caption{Variation of $E(B-V)$ with the spectral class of the stars.
    The gray open circles represent the data points while the black solid points are
   mean $E(B-V)$.} 
    \label{sp}
  \end{figure}

To understand the relation between the interstellar extinction and the spectral type of the
stars, we plotted $E(B-V)$ against the spectral type of the stars with gray open points 
in Fig. \ref{sp}. 
To see the significant variations, we have binned
the stars in groups and calculated mean and standard deviation of $E(B-V)$. The plots
of mean $E(B-V)$ against spectral type are shown in Fig. \ref{sp} with black solid points.
These plots clearly show that for most of the clusters there is no significant trend in $E(B-V)$
with spectral class except Berkeley 7, Hogg 10, NGC 6871 and NGC 7380.
For the clusters NGC 7380, NGC 6871 and Berkeley 7 initially $E(B-V)$ remains constant, but
after $B6$ spectral class, it decreases. In Hogg 10, $E(B-V)$ decreases continuously from $O$
to $A$ type stars.


\begin{figure}
    \centering
    \includegraphics[width=9cm,height=8cm]{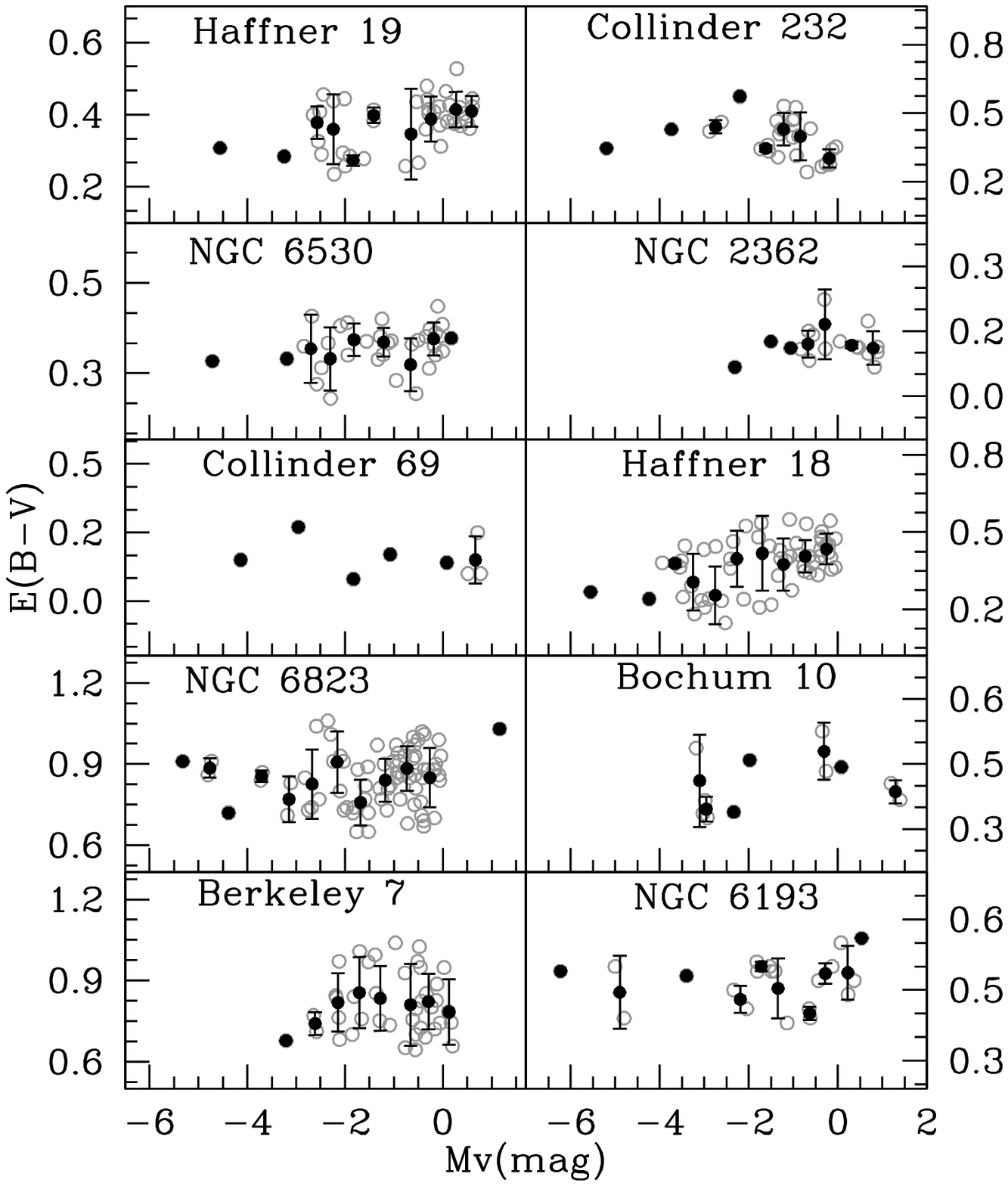}
    \includegraphics[width=9cm,height=8cm]{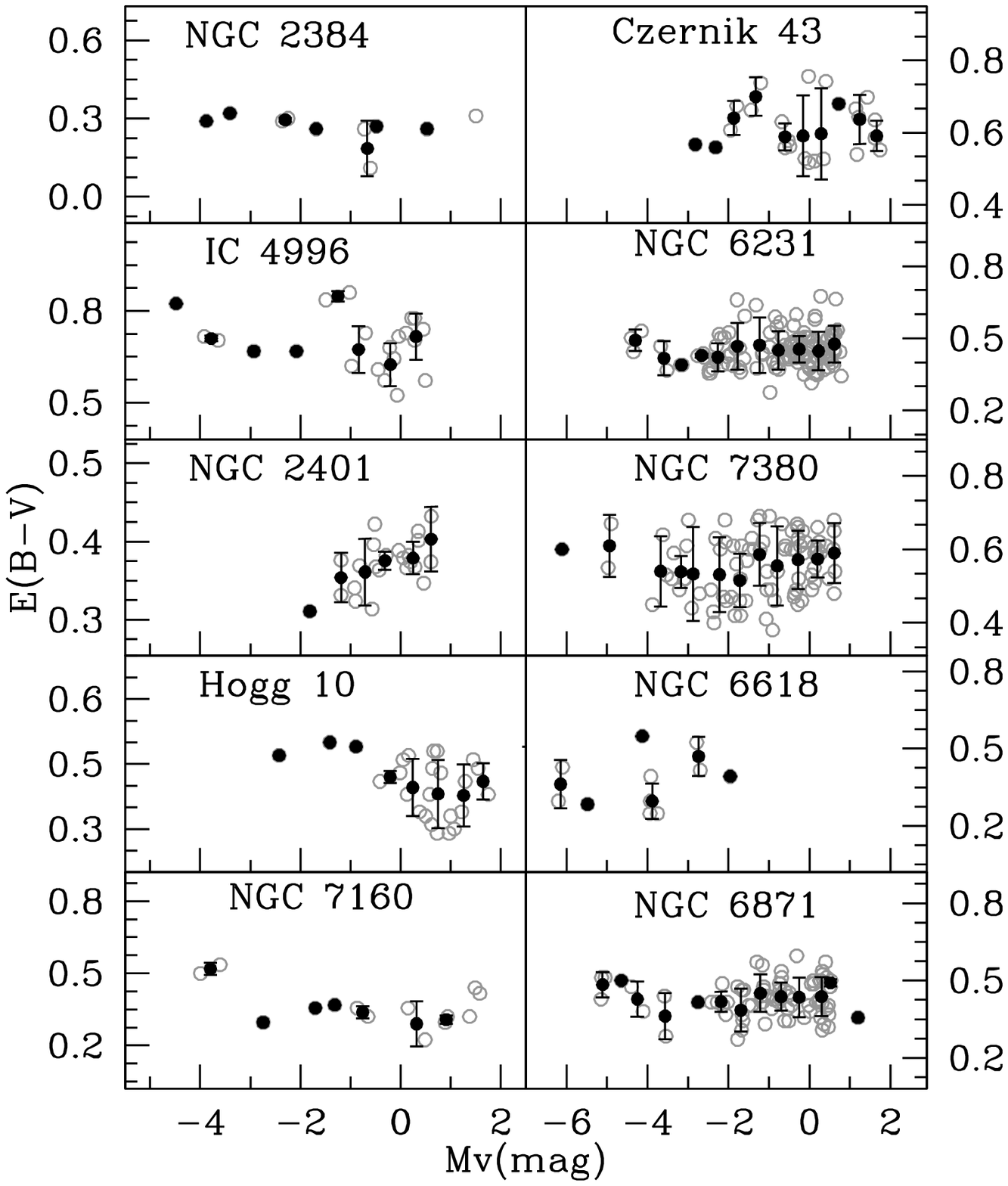}
    \caption{Variation of $E(B-V)$ with the luminosity of the stars.
    The gray open circles represent the data points while the solid black points are
    mean $E(B-V)$}. 
    \label{luminosity}
  \end{figure}

\subsection{Colour excess variation with luminosity} \label{sec:lum}

To study the variation of $E(B-V)$ with the luminosity of stars, we have plotted the
colour excess $E(B-V)$ against the absolute magnitude of the stars. These plots are shown
in Fig. \ref{luminosity} with gray open points. Apparent $V$ magnitudes are converted into absolute magnitude
$M_{V}$ by the relation $M_{V}=V-$(5logd$-5+A_{V})$ where 
$A_{V}=3.25E(B-V)$ and distances are taken from Table \ref{inf}.
To see the clear variation in the colour excess, we have plotted the mean and standard
deviation of $E(B-V)$ against the absolute luminosity and shown in Fig. \ref{luminosity} with the black solid points with errors.
We grouped the stars to calculate the mean and standard deviation in colour excess.
A close inspection
of Fig. \ref{luminosity} shows that there is no significant correlation of mean $E(B-V)$
with luminosity for clusters Collinder 69, NGC 2384, NGC 6618, NGC 7160, Haffner 18, Collinder 232,
Czernik 43, NGC 2362, NGC 2401, Haffner 19, NGC 6823, NGC 2362, IC 4996, NGC 6231 and NGC 6618.
However, in the clusters NGC 7380, NGC 6871 and Berkeley 7 a decreasing trend of 
$E(B-V)$ with luminosity is observed after $M_v$$\sim$$0$ mag. In Hogg 10, $E(B-V)$ decreases
continuously with $M_v$.

\subsection{Discussions of variable reddening} \label{dis}

On the basis of the study presented in sections \ref{sec:sprl} and \ref{sec:lum}, we divided
the clusters sample into following three groups:

1) Clusters namely Berkeley 7, NGC 7380, NGC 6871 belong to group 1 show a dependence
of $E(B-V)$ on both spectral type and luminosity. Initially, $E(B-V)$ remains
constant up to $B6$ spectral class and $M_{v} \sim$ 0 mag luminosity but decreases as we go
towards late type spectral class and fainter stars.

2) For the cluster Hogg 10 which is in group 2, $E(B-V)$ continuously decreases with
luminosity and temperature of the stars.

3) The clusters belonging to group 3 are Collinder 69, NGC 2384, NGC 6618, NGC 7160, Haffner 18,
Collinder 232, Czernik 43, NGC 2362, NGC 2401, Haffner 19, NGC 6823, NGC 2362,
IC 4996, NGC 6231, Hogg 10 and NGC 6618. For these clusters no dependence of $E(B-V)$ on either
spectral type or luminosity is seen.

The observed dependence of $E(B-V)$ on the spectral type and luminosity
of the stars in group 1 and 2 is due to the presence of circumstellar matter around them as
discussed by Sagar (1987). In these groups the value of $E(B-V)$ is higher
for early type and most luminous stars. Yorke \& Krugel (1977) and Bhattacharjee \&
Williams (1980) discussed that relative mass of the relict envelope present
around a newly formed massive star is positively correlated with the stellar mass.
Massive main sequence stars blow off their circumstellar material by emitting strong stellar
winds and ultraviolet radiations. But the time required to blow off this circumstellar material
is dependent on various parameters as discussed by Sagar (1987). Therefore, stars of group
1 and 2 have their circumstellar material still present around them whereas group
3 stars have blown their circumstellar envelope away.

\subsection{Near-IR fluxes} \label{ir}

The total interstellar extinction towards a star is the
summation of three terms as stated by Yadav \& Sagar (2001). These three terms are:
(i) interstellar extinction caused by the dust and gas present between the observer and
the cluster. (ii) intra-cluster extinction caused by the dust and gas present inside the
cluster field and (iii) circumstellar extinction caused by the remnant discs or envelope
around the young cluster members in which the disc is not completely dissipated.

\begin{figure}
    \centering
    \includegraphics[width=9cm,height=9cm]{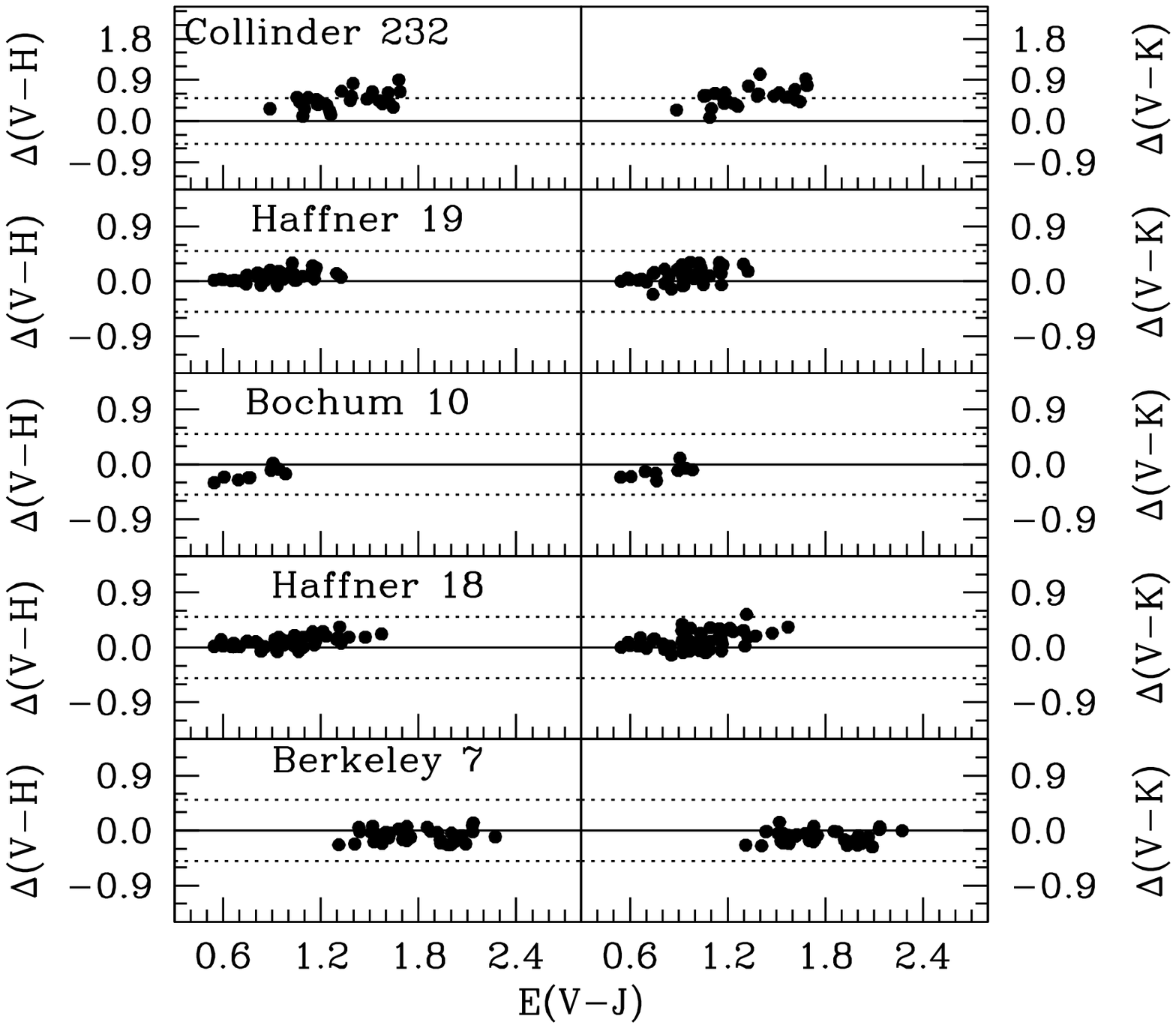}
    \includegraphics[width=9cm,height=9cm]{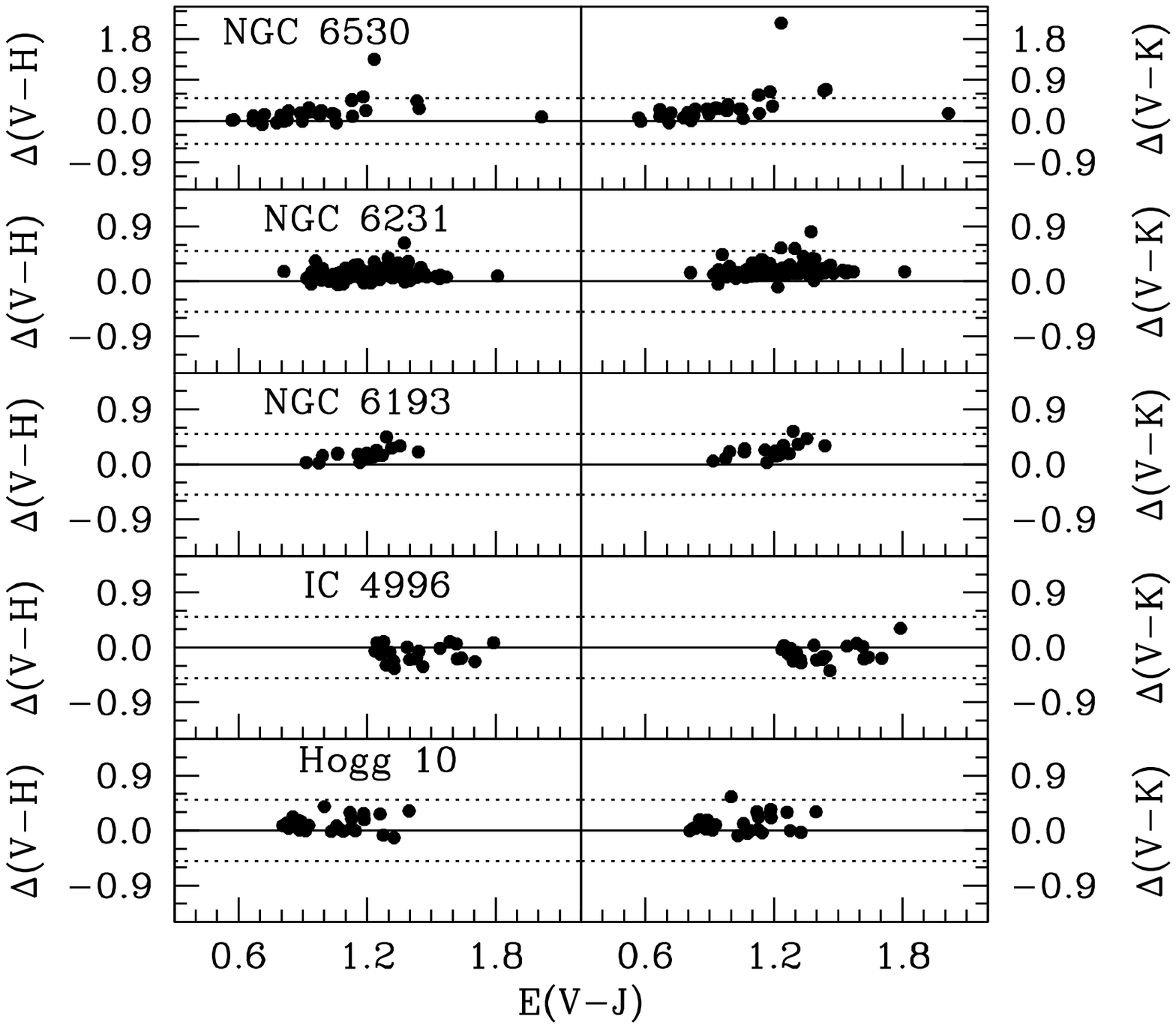}
    \caption{Plots of near-IR flux excess in terms of $\Delta(V-H)$ and
$\Delta(V-K)$ against the colour excess $E(V-J)$. The black line
denotes the zero excess and the broken lines denote the extent of significant error.}
    \label{fl1}
  \end{figure}

\begin{figure}
    \centering
    \includegraphics[width=9cm,height=9cm]{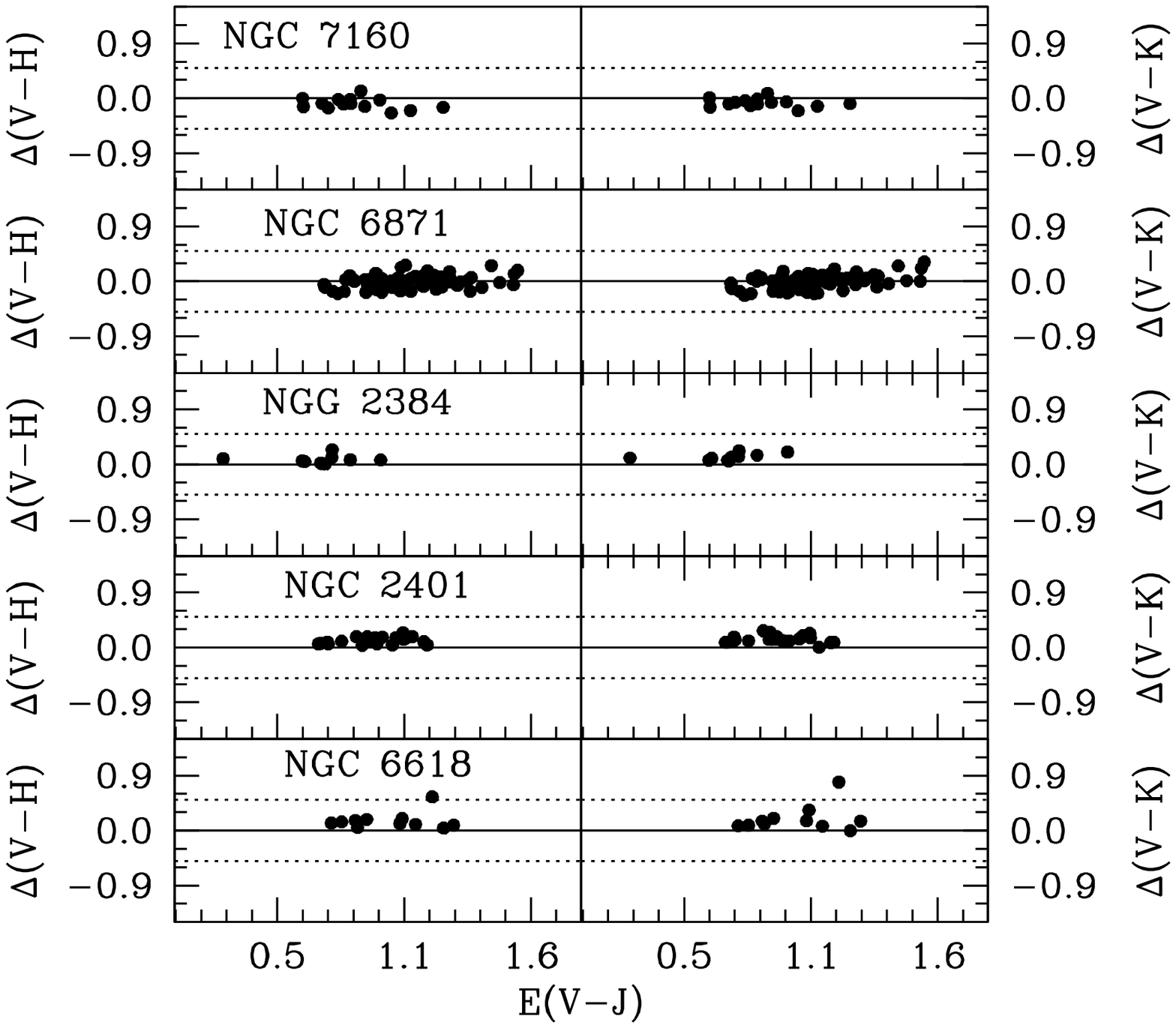}
    \includegraphics[width=9cm,height=9cm]{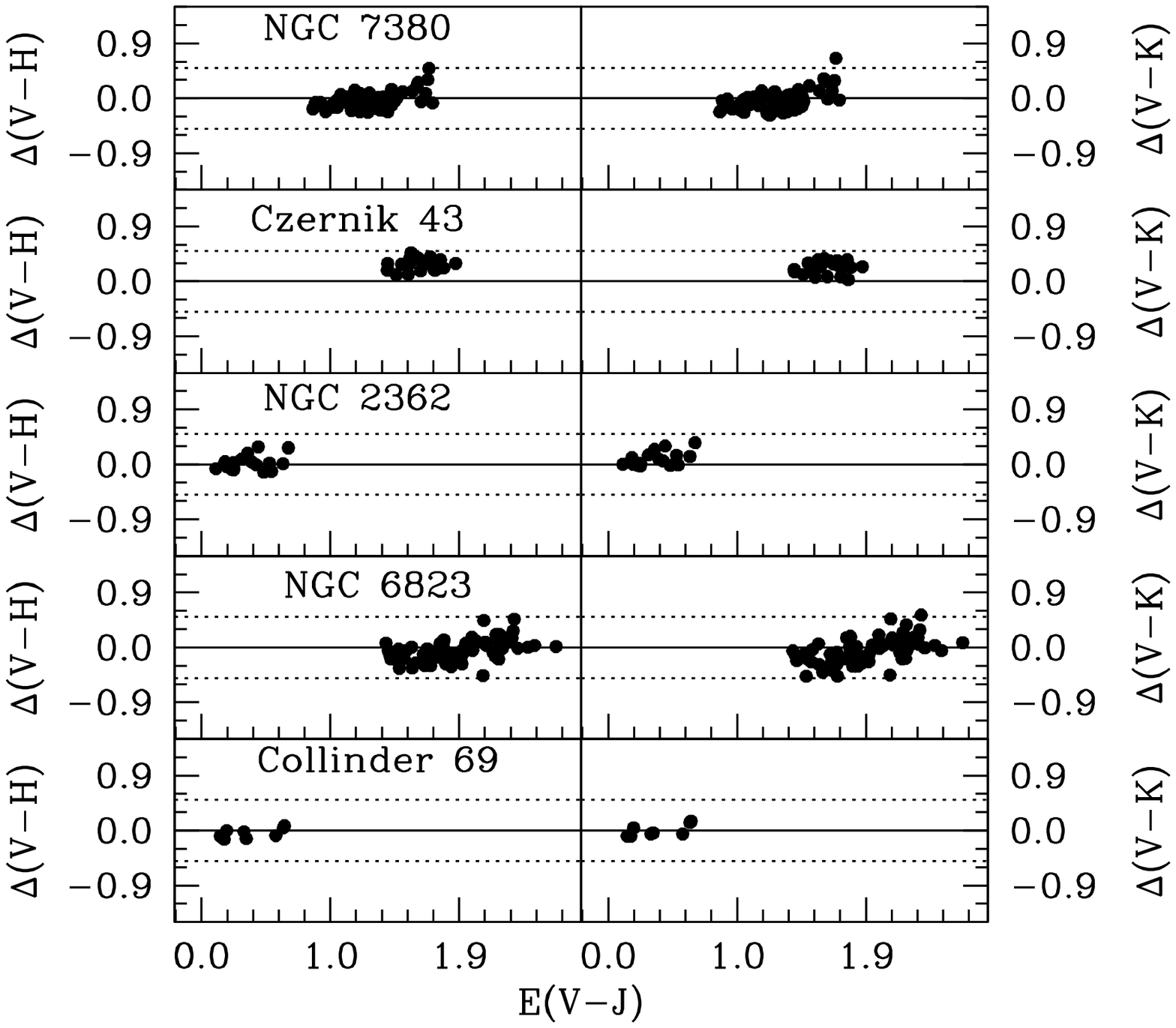}
    \caption{Plots of near-IR flux excess in terms of $\Delta(V-H)$ and $\Delta(V-K)$
against the colour excess $E(V-J)$. The black line
denotes the zero excess and the broken lines denote the extent of significant error.}
    \label{fl2}
  \end{figure}

Generally it is assumed that the interstellar extinction caused by general interstellar medium
between observer and the object follows the normal interstellar extinction law. So
if there is an anomaly in the observed extinction law, then it may be due to the intra-
cluster extinction or due to the circumstellar extinction or due to both. Since in the
present study we have selected brighter stars in young clusters so
it may be possible that the observed anomaly in the interstellar extinction law may be due
to the reasons listed in points (ii) and (iii) in above paragraph. To test these two possibilities we have to use the
colour excess $E(V-J)$ instead of $E(B-V)$ as stated by Smith (1987), Tapia et al. (1988),
Sagar \& Qian (1989, 1990) and Yadav \& Sagar (2001) that the value of $E(V-J)$
is independent of the properties such as chemical composition, shape, structure and
degree of alignment of the interstellar dust and cloud (Vashchinnikov \& I1'in 1987
; Cardelli et al. 1989). In order to study the reason behind the
anomaly in the observed interstellar law, we have calculated the differences between
the observed values of $E(V-H)$ and $E(V-K)$ based on spectral classification and the
derived colour excess values from $E(V-J)$ assuming normal extinction law
and has been plotted against the corresponding value of $E(V-J)$ in Fig. \ref{fl1} and
\ref{fl2}.

\begin{table}
\tiny
\centering
\caption{Near-IR flux excess in stars of cluster under study. Star numbers are taken from WEBDA}
\begin{tabular}{ccccccc}
\hline\hline
Cluster    &  Stars    &  $V$  &  $E(V-J)$  &  $\Delta(V-H)$  &  $\Delta(V-K)$ & SP \\
           &           & (mag) & (mag) & (mag) & (mag)  &  type   \\
\hline
Collinder 232       &    3 &  12.66 &  1.18  &  0.35 & 0.61  & B3  \\
               &    4 &  14.06 &  1.12  &  0.51 & 0.60  & B9.5  \\
               &   27 &  14.19 &  1.39  &  0.82 & 1.02  & B9.5  \\
               &   41 &  13.22 &  1.68  &  0.90 & 0.92  & B8   \\
               &   46 &  13.62 &  1.51  &  0.64 & 0.61  & B8   \\
               & 1011 &  11.62 &  1.61  &  0.62 & 0.69  & B3   \\
               & 1025 &  13.02 &  1.68  &  0.64 & 0.77  & B3   \\
               & 1050 &  14.11 &  1.32  &  0.65 & 0.76  & B9   \\
NGC 6530       &  26  &  11.59 &  1.44  &  0.27 & 0.69  & B5   \\
               &   31 &  11.69 &  1.18  &  0.53 & 0.64  & B6   \\
               &   64 &  11.63 &  0.04  &  0.81 & 2.08  & B6   \\
               &   263&  11.43 &  1.43  &  0.44 & 0.66  & B5   \\
NGC 6231       &  370 &  12.23 &  1.37  &  0.62 & 0.81  & B8   \\
\hline
\end{tabular}
\label{fl3}
\end{table}

In the Fig. \ref{fl1} and \ref{fl2}, $\Delta(V-H)$ and $\Delta(V-K)$ are the differences
of observed and derived colour excess, the continuous line denote the zero value of $\Delta(V-H)$
and $\Delta(V-K)$.
The errors in $\Delta(V-H)$ and $\Delta(V-K)$ mainly occur due to observational uncertainties in $JHK$ magnitudes,
inaccuracies in the calculation of $E(V-J)$ and errors in spectral classification.
The maximum value of differences due to these factors can be $\sim 0.5$ mag and denoted
by dotted lines in Fig. \ref{fl1}.

On close inspection of Fig.
\ref{fl1} and \ref{fl2}, we can see that for most of the clusters, stars are concentrated
around the zero value and inside the error line, which indicates that there is no near-IR
excess fluxes in the stars of the clusters. In the clusters namely Collinder 232, NGC 6530 and NGC 6231
there are some stars which have larger value of $\Delta(V-H)$ and $\Delta(V-K)$
and they are lying outside the error boundary. These stars are listed in Table~\ref{fl3}.

All the stars listed in Table~\ref{fl3} for the clusters Coll 232, NGC 6530 and NGC 6231
have larger values in $\Delta(V-K)$. It is possible that circumstellar material is
present around these stars. We have detected near-IR fluxes for these stars
for the first time. Strom et al. (1971, 1971) noticed the presence of circumstellar
material around NGC 2264-90 on the basis of near-IR fluxes. Smith (1987) and Yadav \&
Sagar (2001) have also found near-IR excess fluxes for the stars Tr 14-15, NGC 2264-165 and Tr 16-68.

\section{Conclusions} \label{con}

In the present study we analyzed the interstellar extinction law in the direction of twenty
open clusters younger than 50 Myr. We have also studied the variation of colour excess $E(B-V)$
with age, luminosity, spectral type and position of stars in the cluster. The main results of
this study are summarized as follows.

\begin{enumerate}

\item The histograms of $E(B-V)$ show that gas and dust is non-uniformly distributed over the
cluster area in each cluster. The value of $\Delta E(B-V)$ confirms the presence of non-uniform
extinction in all clusters under study except NGC 2362, NGC 2384 and Collinder 69. Our analysis
shows that non-uniform extinction decreases with the age of the clusters. This indicates that
dust and gas are still present in younger clusters while it has been blown away in older clusters
by UV radiations of hot stars.

\item Derived colour excess ratios show that twelve clusters NGC 6823, Haffner 18, Haffner 19,
NGC 7160, NGC 6193, NGC 2401, NGC 2384, NGC 6871, NGC 7380, Berkeley 7, Collinder 69 and IC 4996
have normal extinction laws in the optical and near-IR regions within $2\sigma$ error. This indicates
that dust and gas present in the clusters have normal size distributions. Six clusters
Bochum 10, NGC 2362, Collinder 232, Hogg 10, NGC 6618 and NGC 6231 show anomalous behaviour in
the optical band. Two clusters NGC 6530 and Czernik 43 show anomalous behaviour for
$\lambda \geq \lambda_{J}$. This shows that the extinction law for these clusters is different from normal one.

\item Spatial variation of the colour excess $E(B-V)$ has been found only for one
cluster, NGC 6193, in the sense that $E(B-V)$ is decreasing from east to west. For other
clusters, $E(B-V)$ is distributed non-uniformly over the cluster field.

\item A small variation in $E(B-V)$ with spectral class and luminosity of stars has been
found in the clusters Berkeley 7, NGC 7380, NGC 6871 and Hogg 10. The observed dependency of
$E(B-V)$ shows that early type stars in these clusters are having some amount of dust and
gas around them.

\item From the near-IR flux values, we identified eight stars in Collinder 232,
four stars in NGC 6530 and one star in NGC 6231 which are having significant
near-IR excess fluxes. Higher fluxes may be due to the presence of circumstellar material
around these stars.

\end{enumerate}

{\bf ACKNOWLEDGMENTS}\\

We are thankful to the anonymous referee for carefull reading of the paper 
and constructive comments.
We also thank Prof. Johan P. U. Fynbo, Niels Bohr Institute, University of Copenhagen,
Denmark for reading our manuscript. 
This research has made use of the WEBDA database, operated at the Department of
Theoretical Physics and Astrophysics of the Masaryk University. We have also
used data from Two Micron All Sky Survey, which is a joint project of the
University of Massachusetts and the Infrared Processing and
Analysis Center/California Institute of Technology, funded
by the National Aeronautics and Space Administration and
the National Science Foundation.


\end{document}